\documentclass[a4paper,12pt]{article}
\usepackage{jcappub} % for details on the use of the package, please
\usepackage{hyperref}
\hypersetup{
    bookmarks=true,         % show bookmarks bar?
    unicode=false,          % non-Latin characters in Acrobat’s bookmarks
    pdftoolbar=true,        % show Acrobat’s toolbar?
    pdfmenubar=true,        % show Acrobat’s menu?
    pdffitwindow=false,     % window fit to page when opened
    pdfstartview={FitH},    % fits the width of the page to the window
    pdftitle={My title},    % title
    pdfauthor={Author},     % author
    pdfsubject={Subject},   % subject of the document
    pdfcreator={Creator},   % creator of the document
    pdfproducer={Producer}, % producer of the document
    pdfkeywords={keyword1} {key2} {key3}, % list of keywords
    pdfnewwindow=true,      % links in new window
    colorlinks=true,       % false: boxed links; true: colored links
    linkcolor=red,          % color of internal links
    citecolor=cyan,        % color of links to bibliography
    filecolor=magenta,      % color of file links
    urlcolor=blue,           % color of external links
    linktocpage=true
}
\usepackage{subfigure}
\usepackage{pstricks}
\usepackage{color}
\usepackage{amsfonts}
\usepackage{mathrsfs}
\usepackage{epsfig}
\usepackage{amsmath,amssymb,amsthm,graphicx,latexsym}
\usepackage{ulem}

\newcommand{\be}{\begin{equation}}
\newcommand{\ee}{\end{equation}}
\newcommand{\bea}{\begin{eqnarray}}
\newcommand{\eea}{\end{eqnarray}}
\newcommand{\bml}{\begin{subequations}}
\newcommand{\eml}{\end{subequations}}
\newcommand{\bfig}{\begin{figure}}
\newcommand{\efig}{\end{figure}}

\newcommand{\slashed}{\hspace{-1.1ex}/}

\title{Constraining  ${\cal N}=1$ supergravity inflationary framework with non-minimal K\"ahler operators}

\author[a]{ Sayantan Choudhury}
\author[b]{ Anupam Mazumdar}
\author[b]{ Ernestas Pukartas}

\affiliation[a]{ Physics and Applied Mathematics Unit, Indian Statistical Institute, 203 B.T. Road, Kolkata 700 108, INDIA}
\affiliation[b]{ Consortium for Fundamental Physics, Physics Department, Lancaster University, LA1 4YB, UK}

\abstract{In this paper we will illustrate how to constrain unavoidable K\"ahler corrections for ${\cal N}=1$ supergravity (SUGRA)
inflation from the recent Planck data. We will show that the non-renormalizable K\"ahler  operators will induce in general {\it non-minimal kinetic} term for the inflaton field,
and two types of SUGRA corrections in the potential - the {\it Hubble-induced mass} ($c_{H}$), and the {\it Hubble-induced A-term} ($a_{H}$) correction.
The entire SUGRA inflationary framework can now be constrained from (i) the {\it speed of sound}, $c_s$, and (ii) from the upper bound on the {\it tensor to scalar ratio}, $r_{\star}$. 
We will illustrate this by considering  a heavy scalar degree of freedom at a scale, $M_s$, and a  light inflationary field which is responsible for a slow-roll inflation. We will compute 
the corrections to the kinetic term and the potential for the light field explicitly. As an example, we will consider a visible sector inflationary model of inflation where inflation occurs at the point 
of {\it inflection}, which can match 
the density perturbations for the cosmic microwave background radiation, and also explain why the universe is filled with the Standard Model degrees of freedom. 
We will scan the parameter space  of the non-renormalizable K\"ahler operators, which we find them to be order ${\cal O}(1)$, consistent with physical arguments.
While the scale of heavy physics is found to be bounded by the tensor-to scalar ratio, and the speed of sound,  $ {\cal O}(10^{11}\leq M_s\leq 10^{16})$~GeV, for
$0.02\leq c_s\leq 1$ and $10^{-22}\leq r_\star \leq 0.12$. }

\begin{document} 
\maketitle
\flushbottom

\section{Introduction}

The success of primordial inflation~\cite{Guth:1980zm}, for a review, see~\cite{Mazumdar:2010sa}, can be gauged by the current observations arising from the comic microwave background (CMB) radiation~\cite{Planck-infl,Planck-1,Ade:2013ydc}. The observations from Planck have put interestingly tight bounds on a number of unknown parameters of a generic inflationary model~\cite{Planck-infl}, in particular the {\it speed of sound}, $c_s$, of the perturbations, which also determines any departure from the Gaussian perturbations, the local type of non-Gaussianity, $f_{NL}^{\mathrm{local}}$,  and the constraint on {\it tensor-to-scalar ratio}, $r_\star$, which can potentially unearth the scale of New Physics within any given {\it effective field theory } set-up.

The ${\cal N}=1$ supergravity (SUGRA)~\cite{Freedman:1976xh}, for a review, see~\cite{Nilles:1983ge}, is an excellent well-defined set-up where we can address some of the key questions about the physics of the new scale for instance. The K\"ahler metric determines the kinetic term for the inflaton  potential, and one particular choice is the {\it minimal} Kinetic term for the inflaton field. However, quantum corrections to the K\"ahler potential is not very well-known. Generically corrections to the K\"ahler potential arise from integrating out the heavy physics, and due to lack of concrete knowledge on the details of the heavy physics, many times computing these corrections can be very challenging, see~\cite{Berg:2005ja}. 

The aim of this paper is to place a generic bound on the Planck suppressed corrections to the K\"ahler potential on top of the minimal K\"ahler potential. We will consider dimensional $3$ and dimensional $4$ {\it gauge invariant} non-renormalizable K\"ahler operators in this paper. Since these correction will lead to a departure from the minimal kinetic term for the inflationary potential, such corrections can now be bounded from the Planck data, especially from the speed of sound of the primordial perturbations. In fact the K\"ahler potential within ${\cal N}=1$ SUGRA can induce corrections to the inflationary potential which can yield large Hubble-induced mass correction to the inflaton field~\cite{Bertolami:1987xb,Mazumdar:2011ih}, some times known as the SUGRA-$\eta$ problem, and the Hubble-induced SUGRA $A$-term for the potential.  In the context of ${\cal N}=1$ SUGRA hybrid inflation~\cite{Rehman:2010wm,Khalil:2010cp,Shafi:2010jr}, some of the K\"ahler corrections were constrained by the tensor to scalar ratio, $r_\star$, and the spectral tilt, $n_S$, of the power spectrum. In this paper we will be constraining these K\"ahler corrections systematically from the interaction of the heavy physics to the light inflaton field from the recent Planck data.
Let us consider the scale of New Physics, $M_s$, to be within
\begin{equation}
 M_p\geq M_s\geq H_{inf}\,,
 \end{equation}
where $H_{inf}$ is the Hubble parameter during inflation, and $M_p=2.4\times 10^{18}$~GeV.
In order to constrain $M_s$, we would require to consider at least $2$ fields, one which is heavy at the relevant scale, $M_s$,  and the other which 
is light. We will assume that these two fields are coupled gravitationally. In past such a scenario has been considered by many authors, 
where the heavy field leaves interesting imprints in the dynamics of a low scale inflation~\cite{Burgess:2002ub,Burgess:2003zw,Burgess:2012dz,Achucarro:2012yr,Achucarro:2012sm,Collins:2009pf,Achucarro:2010da,Achucarro:2010jv,Assassi:2013gxa,Agarwal:2013rva,Antusch:2008pn}. Broadly speaking there are two possible scenarios which one can envisage:

\begin{itemize}

\item {\bf The heavy field is dynamically frozen}: we can imagine that the heavy field is completely {\it frozen}, in which case 
it would be effectively a single light field with  a canonical kinetic term for the inflaton field, with a speed of sound, $c_s=1$. 
For a slow roll inflation, the perturbations will be primarily Gaussian. If the heavy field is settled down to its minimum VEV, i.e. zero, then there will be no effect from 
the heavy field at all. However, if the heavy field is settled with a finite non-zero VEV, and it remains dynamically inactive, means its velocity is strictly zero, then
it can still contribute to the vacuum energy density of the inflaton, and this would be encoded in $H_{inf}$. Also, the kinetic term for the light inflaton field will 
depart from being pure canonical. However the departure will depend on the scale of new physics. If $M_s\ll M_p$, then the departure from canonical kinetic term
will be negligible for all practical purposes. Therefore, again the observational predictions for the CMB will be unaltered and will be similar to the previous case. 
Both of these scenarios were taken into account by various interesting papers, see for example~\cite{Rehman:2010wm,Shafi:2010jr,Mazumdar:2011ih,Antusch:2010mv,Antusch:2012jc,Choudhury:2013jya}, and here we will not consider them in great details. We will analyse a slightly different scenario as mentioned below.

\item{\bf The heavy field is coherently oscillating during the initial phases of  inflation}:  in this case we will consider a very simple scenario, where we imagine that the heavy field is coherently oscillating at a VEV given by $M_s$ with an amplitude $M_s$ at the onset of inflation driven by the light field. The coherent oscillations of the heavy field will not last forever, its amplitude would be damped during inflation very rapidly within couple of e-foldings of inflation. However, just right at the onset of inflation, the relevant modes which are leaving the Hubble patch for the CMB can be constrainable. This will provide a window of opportunity for us to constrain such a scenario, see Refs.~\cite{Burgess:2002ub,Burgess:2003zw,Burgess:2012dz} for probing the influence of heavy physics into the light inflaton field. In this paper we will consider a similar scenario, but in the context of ${\cal N}=1$ SUGRA.

First of all the coherent oscillations of the heavy field around its non-zero vacuum would provide a non-zero vacuum energy density, i.e $\sim M_s^4$. Through its coupling to the light field in the K\"ahler potential, it would also yield non-canonical kinetic term contribution to the light field, and therefore $c_s\neq 1$ for a slow roll inflaton field. Eventually, the heavy field will be settled down to its minimum. We presume that the dominant contribution to the long wavelength fluctuations are still seeded by the light  inflaton, but the fact that  $c_s\neq1$ for the inflaton, it would leave imprints which would be constrainable directly by the scale of heavy physics, $M_s$, and the non-minimal corrections to the k\"ahler potential.
\end{itemize}

We will discuss this latter scenario in some details, and provide a full ${\cal N}=1$ SUGRA potential for the light and the heavy field within a simple example.
 We will be using the following constraints from CMB, and also requirement for a guaranteed reheating of the Standard Model d.o.f for the success of 
 big bang nucleosynthesis~\cite{Fields:2006ga}.

\begin{enumerate}

\item{Successful single field inflation driven by $\phi$ field with the right amplitude and tilt of the power spectrum.
\begin{eqnarray}\label{powq1}
\displaystyle 2.092<10^{9}P_{S}<2.297~~(~within~2\sigma),\\
\label{powq2}
\displaystyle 0.958<n_{S}<0.963~~(~within~2\sigma)\,.
 \end{eqnarray}
}

\item{Speed of sound, $c_s$: The Planck analysis has constrained it to be~\cite{Planck-infl,Ade:2013ydc}:
\begin{equation}\label{c-s}
\displaystyle 0.02 \leq c_s \leq 1~~(~within~2\sigma)\,.
\end{equation}
}
\item{Tensor-to-scalar ratio: The Planck constraint is $r_\star\leq 0.12$~\cite{Planck-infl}
\begin{equation}\label{t-s}
\displaystyle r_\star \equiv \frac{P_T}{P_S}\leq 0.12.
\end{equation}
}

\item{Local type of non-Gaussianity, $f_{NL}^{\mathrm{local}}$: The Planck constraint on local non-Gaussianity is~\cite{Planck-infl,Ade:2013ydc}:
%~\footnote{When $c_s\neq 1$, we would also generate primordial non-Gaussianity of order $f_{NL}\propto {\cal O}(1/c^{2}_{s})$.
 %} :
\begin{eqnarray}\label{f-nl}
\displaystyle f_{NL}^{\mathrm{local}}=2.7\pm 5.8~~(~within~1\sigma)\,.
\end{eqnarray}
In this paper we will not consider the constraints arising from various non-Gausisnaity bounds~\cite{Ade:2013ydc},
but we will solely focus  on the constraints arising from the speed of sound during perturbation, and the tensor to scalar ratio.
In the companion paper we have discussed how non-Gaussianity can constrain the K\"ahler corrections.}

\item{Particle physics constraint: We wish to ensure that the inflaton {\it solely} decays into the Standard Model (SM) 
d.o.f, therefore we embed the light fields within supersymmetric SM, such as minimal supersymmetric Standard Model (MSSM).
In this case the inflaton carries {\it solely} the SM charges as in the case of 
the inflaton driven by {\it gauge invariant} combinations of squarks and sleptons~\cite{Allahverdi:2006we,Allahverdi:2006iq,Allahverdi:2006cx,Chatterjee:2011qr}. 
This will naturally ensure that we obtain the right abundances for the dark matter and the baryons in the universe as required by the observations~\cite{Allahverdi:2007vy,Balazs:2013cia}. One can however follow a hidden sector or a SM gauge singlet inflaton, but it is not always straightforward to explain the universe with the right dark matter
abundance~\cite{Dev:2013yza}, and baryon asymmetry, see~\cite{Mazumdar:2010sa}.}

\end{enumerate}

The results of the first half of this paper will be very generic - applicable to any inflationary scenario.
In section \ref{perturb}, we will discuss briefly the Planck constraints. In section \ref{setup}, we will discuss the setup with one heavy and one light superfield which are coupled via gravitational interactions through
 K\"ahler potential. In section \ref{eft}, we will describe the effective field theory potential for the light superfield $\Phi$, and discuss the kinetic terms for 
 various interesting scenarios. In section \ref{N-M}, we will discuss the role of non-canonical kinetic term and consider two possibilities, one where the heavy superfield
 is dynamically frozen, see section \ref{frozen}, and the more interesting scenario when the heavy field is oscillating at the onset of inflation, see section~\ref{oscillating}. We 
will  scan the parameters for the Planck suppressed K\"ahler operators in subsection~\ref{02},  we will discuss how tensor-to-scalar ratio, $r_\star$, can constrain the mass 
scale of the heavy physics.

%%%%%%%%%%%%%%%%%%%%%%%%%%%%%%%%%%%%%%%%%%%%%%%%%%
\section{ Cosmological perturbations for $c_s\neq 1$}\label{perturb}

In this section we briefly recall some of the important formulae when $c_s\neq 1$, the scalar and tensor perturbations are given by~\cite{Garriga:1999vw,Mukhanov:1990me,Planck-infl}:
\begin{eqnarray}\label{scalaras}
P_{S}(k)& = &{\cal P_{S}}\left(\frac{k}{c_{s}k_{\star}}\right)^{n_{S}-1}\,,\nonumber \\
\label{tensoras}P_{T}(k)& = &{\cal P_{T}}\left(\frac{k}{c_{s}k_{\star}}\right)^{n_{T}},
\end{eqnarray}
where the speed of sound at the Hubble patch is given by, $c_{s}k_{\star}=aH$ (where
$k_{\star}\sim 0.002~Mpc^{-1}$). The amplitude of the scalar and tensor perturbations can be recast in terms of the potential, as~\cite{Planck-infl}:
\begin{eqnarray}\label{scalar}
{\cal P_{S}}& = &\frac{V_{\star}}{24\pi^{2}M^{4}_{p}c_{s}\epsilon_{V}}\,, \\
\label{tensor}{\cal P_{T}}& = &\frac{2V_{\star}}{3\pi^{2}M^{4}_{p}}c^{\frac{2\epsilon_{V}}{1-\epsilon_{V}}}_{s},
\end{eqnarray}
where running of the spectral tilt for the scalar and tensor modes can be expressed 
at $c_{s}k_{\star}=aH$, as:
\begin{eqnarray}\label{scalartilt}
n_{S}-1& = &2\eta_{V}-6\epsilon_{V}-s\,, \\
\label{tensortilt}n_{T}& = &-2\epsilon_{V}.
\end{eqnarray}
where running of the sound speed is
defined by  an additional slow-roll parameter, $s$, as:
\be\begin{array}{llll}\label{cvq1}
    \displaystyle s=\frac{\dot{c_{s}}}{Hc_{s}}=\sqrt{\frac{3}{V}}\frac{\dot{c_{s}}}{c_{s}}M_{p}.
   \end{array}\ee
In all the above expressions, the standard slow-roll parameters (with $c_s=1$) are defined by:
\be\begin{array}{llll}\label{cvq1}
    \displaystyle \epsilon_{V}=\frac{M^{2}_{p}}{2}\left(\frac{V^{\prime}}{V}\right)^{2}, ~~~~~
     \eta_{V}={M^{2}_{p}}\left(\frac{V^{\prime\prime}}{V}\right).
   \end{array}\ee
Finally, the single field consistency relation between tensor-to-scalar ratio
 and tensor spectral tilt is modified by~\cite{Planck-infl,Powell:2008bi}:
\be\begin{array}{llll}\label{sconfo}
    \displaystyle r_{\star}=16\epsilon_{V}c^{\frac{1+\epsilon_{V}}{1-\epsilon_{V}}}_{s}=-8n_{T}c^{\frac{1-\frac{n_{T}}{2}}{1+\frac{n_{T}}{2}}}_{s}.
   \end{array}\ee
 Using the results for $c_{s}\neq 1$ stated in Eqs.~(\ref{scalar}-\ref{sconfo}), the upper bound on the numerical value of the Hubble 
 parameter during inflation is given by:
\begin{equation}\label{hubble}
     H\leq 9.241\times 10^{13}\sqrt{\frac{r_{\star}}{0.12}}~c^{\frac{\epsilon_{V}}{\epsilon_{V}-1}}_{s}~{\rm GeV}\,
   \end{equation}
where $r_{\star}$ is the tensor-to-scalar ratio at the pivot scale of momentum $k_{\star}\sim 0.002 Mpc^{-1}$.
An equivalent statement can be made in terms of the upper bound
on the energy scale of  inflation for $c_{s}\neq 1$ as:
\begin{equation}\label{scale}
     V_{\star}\leq (1.96\times 10^{16}{\rm GeV})^{4}\frac{r_{\star}}{0.12}~c^{\frac{2\epsilon_{V}}{\epsilon_{V}-1}}_{s}.
   \end{equation}
Here in Eqs.~(\ref{hubble}) and (\ref{scale}), the equalities will hold good for a high scale model of inflation.

Furthermore, for a sub-Plancikan slow-roll models of inflation, one can express the tensor-to-scalar ratio, $r_{\star}$, at the
pivot scale, $k_{\star}\sim 0.002~Mpc^{-1}$, in terms of the field displacement, $\Delta \phi$, during
the observed $\Delta{ N}\approx 17$ e-foldings of inflation, for $c_{s}\neq 1$~\cite{Choudhury:2013iaa,Choudhury:2013woa}:
%\begin{widetext}
\be\begin{array}{llll}\label{con10}
   \displaystyle \frac{3}{25\sqrt{c_{s}}}\sqrt{\frac{r_{\star}}{0.12}}\left|\left\{\frac{3}{400}\left(\frac{r_{\star}}{0.12}\right)-\frac{\eta_{V}(k_{\star})}{2}-\frac{1}{2}
\,\right\}\right|
\displaystyle \approx \frac{\left |\Delta\phi\right|}{M_{p}}\,.
   \end{array}\ee
 where $\Delta\phi=\phi_{cmb}-\phi_{e}<<M_{p}$, where $\phi_{cmb}$ and $\phi_{e}$
 are the values of the inflaton field at the horizon crossing and at the end of inflation.

%%%%%%%%%%%%%%%%%%%%%%%%%%%%%%%%%%%%%%%%%%%%%%%%%%%%%

\section{Inflationary setup within ${\cal N}=1$ SUGRA}\label{setup}

Let us consider two sectors; heavy sector denoted by the superfield $S$, 
and the light sector denoted by $\Phi$.  Let us assume that  the two sectors interact only gravitationally, $S$ could denote the 
hidden sector, while  $\Phi$ could denote the visible sector for example part of MSSM~\cite{Enqvist:2007tf,Lalak:2007rsa}. Note that the origin of $S$ superfield need not be
always hidden sector, within MSSM it is possible to have a false vacuum at very high VEVs, see for instance~\cite{Allahverdi:2008bt}. In the latter case both $S$ and $\Phi$
could be embedded within MSSM for instance~\footnote{By visible sector we mean that the inflaton itself carries the SM charges, such as in the case of MSSM~\cite{Allahverdi:2006we,Allahverdi:2006iq,Allahverdi:2006cx,Chatterjee:2011qr}. In all these examples the inflaton $\Phi$ is the $D$-flat 
direction made up of  squarks and sleptons, see ~\cite{Enqvist:2003gh}, which is lifted by the $F$-term 
of the non-renormalizable superpotential.}. For the purpose of illustration, we will assume $S$ to have a simple superfield potential given by:
\begin{eqnarray}
 W&=&W(\Phi)+W(S)\,,~~~\Phi ={\rm Light}\,,~S={\rm Heavy}\,,\\
& =&\frac{\lambda \Phi^n}{n M_{p}^{n-3}}+\frac{M_s}{2}S^2\,,
\end{eqnarray}
where $n\geq 3$ and $\lambda\sim {\cal O}(1)$, and $\Phi$ superfield is the $D$-flat direction of MSSM.
The scale $M_s$ governs the scale of heavy physics. Furthermore, we will assume $\langle s \rangle ,~\langle \phi\rangle  \leq M_{p}$, 
where both $s$ and $\phi$ are fields corresponding
to the super field $S$ and $\Phi$. There are two flat directions which can drive inflation with $n=6$, which are lifted by 
themselves~\cite{Dine:1995kz}, 
\begin{equation}
\phi =\frac{{\widetilde u}+\widetilde d+\widetilde d}{\sqrt{3}},~~~~~~~~~~~~~~~~~~~~~~\phi=\frac{\widetilde L +\widetilde L+\widetilde e}{\sqrt{3}}
\end{equation}
where $\widetilde u,~\widetilde d$ denote the right handed squarks, and $\widetilde L$ denotes that left handed sleptons and 
$\widetilde e$ denotes the right handed slepton. In this case the inflaton mass for $\phi$ will be given by:
\begin{equation}
m^2_{\phi}=\frac{m^2_{\widetilde L}+m^2_{\widetilde L}+m^2_{\widetilde e}}{3}\,,~~~~~~~~~~~~~~~~~~~~~~~~~
m^2_{\phi}=\frac{m^2_{\widetilde u}+m^2_{\widetilde d}+m^2_{\widetilde d}}{3}\,,
\end{equation}
 for 
$\widetilde L\widetilde L\widetilde e$ and $\widetilde u\widetilde d\widetilde d$ directions respectively. Typically these
masses are set by the scale of SUSY, which is typically of the order of $\geq {\cal O}(1)$~TeV, set by he ATLAS and CMS~\cite{ATLAS,CMS}.

Let us consider minimal K\"ahler potentials for both $\phi $ and $s$. For the purpose of illustration we will consider the simplest choice which produces minimal kinetic term,  and the  corrections are of the form:
\begin{eqnarray}
K=s^{\dagger}s+\phi^{\dagger}\phi +\delta K\,,
\end{eqnarray}
where {\it gauge invariant} K\"ahler corrections:
\begin{eqnarray}
 \label{corrt}
 \delta K=f(\phi^{\dagger}\phi,s^{\dagger}s),~f(s^{\dagger}\phi\phi),~f(s^\dagger s^\dagger\phi\phi),~f(s\phi^{\dagger}\phi)\,.
 \end{eqnarray}
The higher order corrections to the K\"ahler potentials are extremely hard to compute. In the following, we will assume that the leading order corrections 
are of the generic form - allowed by the gauge invariance~\footnote{For MSSM flat directions, some of these corrections were already considered before in the context of Affleck-Dine baryogenesis, see~\cite{Dine:1995kz, Kasuya:2006wf}. For MSSM inflation  this paper is the first to deal with these corrections explicitly.}. For the purpose of illustration, let us consider the following terms:
\begin{eqnarray}~\label{corrt-1}
 K^{(1)} &=& \phi^\dag \phi+s^\dag s+\frac{a}{M_{p}^2}\phi^\dag \phi s^\dag s +\cdots\,, \\
 \label{corrt-2}
 K^{(2)}&=&\phi^\dag\phi +s^\dag s+\frac{b}{2M_{p}}s^\dag\phi \phi + h.c.+\cdots\,, \\
 \label{corrt-3}
 K^{(3)}&=&\phi^\dag \phi+s^\dag s+\frac{c}{4M_{p}^2}s^\dag s^\dag\phi \phi + h.c.+\cdots\,, \\
 \label{corrt-4}
 K^{(4)}&=& \phi^\dag \phi+s^\dag s+\frac{d}{M_{p}}s\phi^\dag \phi + h.c. +\cdots\,,
\end{eqnarray}
where $a,~b,~c,~d$ are dimensionless parameters~\footnote{The $\cdots$ contain higher order terms of type $(1/M_p)^2(\phi^\dag \phi)^2+ (1/M_p)^2(s^\dag s)^2+...$, 
and higher order terms, here we are ignoring them. These corrections have been taken into account in the context of SUGRA hybrid inflation in 
Refs.~\cite{Rehman:2010wm,Khalil:2010cp,Shafi:2010jr,Antusch:2012jc,Antusch:2010mv}. Here we are  mainly interested in considering the effects of heavy field $s$ on the dynamics of a light field $\phi$. }.  These corrections will inevitably lead to a departure 
from the minimal kinetic energy for both the fields. Our aim will be to constrain these unknown parameters, 
i.e. $a,~b,~c,~d$,  and the scale of heavy physics, $M_s$, from the CMB constraints 
mentioned  above in the introduction.

%%%%%%%%%%%%%%%%%%%%%%%%%%%%%%%%%%%%%%%%%%%%%%%%%%%%%%%%%%%%%

\section{Effective field theory potential for inflaton from ${\cal N}=1$ SUGRA}\label{eft}

Typically, the scalar potential in ${\cal N}=1$ SUGRA for the $F$-term can be written in terms of the superpotential, 
$W$, and the K\"ahler potential, $K$, see \cite{Nilles:1983ge}:
%
%\begin{widetext}
\begin{equation}
\label{main}
 V=e^{K(\Phi_i^\dag,\Phi_i)/M_{p}^2}\bigg{[}(D_{\Phi_i}W(\Phi))K^{\Phi_i\bar{\Phi}_j}(D_{\bar{\Phi}_j}W^*(\Phi^\dag))-3\frac{\lvert W(\Phi)\rvert^2}{M_{p}^2}\,,
 \bigg{]}
\end{equation}
%\end{widetext}
%
where $i=\Phi,~S$ in our case, and  $F_\Phi\equiv D_\Phi W=W_\Phi+K_\Phi/M_{p}^2$, and 
$K^{\Phi_i \bar{\Phi}_j}$ is the inverse matrix of $K_{\Phi_i \bar{\Phi}_j}$, and the subscript denotes derivative 
with respect to the field. 

Typically at the leading order, the total potential will get contributions from~\cite{Dine:1995kz}:
\begin{enumerate}
 \item Interaction between flat direction and inflaton via exponential prefactor:
$$e^{K(\phi,\phi^\dagger)/M_p^2} V(s)\,,$$
\item Cross coupling terms between the flat direction induced K\"ahler derivative and the inflaton superpotential: 
$$K_\phi K^{\phi\bar{\phi}}K_{\bar{\phi}} \frac{\left| W(s) \right|^2}{M_p^4}\,,$$
\item Interaction between the K\"ahler derivative and superpotential of the inflaton, supergravity K\"ahler metric and K\"ahler potential of the flat direction:
$$K_\phi K^{\phi\bar{s}} D_{\bar{s}} W^*(s^\dagger) \frac{W(s)}{M_p^2} +h.c.\,,$$
\item Self coupling between inflaton via K\"ahler derivative interaction:
$$\big( D_S W(s) \big) K^{s \bar{s}} \big(D_{\bar{S}} W^*(s^{\dagger}) \big)\,.$$
\end{enumerate}
Additionally, the Hubble-induced A terms arises from the following dominant contributions in the effective theory of 
supergravity ~\cite{Dine:1995kz}: 
\begin{enumerate}
 \item Cross coupling terms in the 
K\"ahler derivative between the derivative of the flat direction superpotential and inflaton 
superpotential:
$$W_\phi K^{\phi\bar{\phi}}K_{\bar{\phi}}\frac{W^*(s^\dagger)}{M_p^2} + h.c.\,,$$
\item Interaction terms between the flat direction superpotential and inflaton K\"ahler derivative:
$$K_s \frac{W(\phi)}{M_p^2} K^{s\bar{s}} \left( D_{\bar{S}}W^*(s^\dagger) \right) + h.c.\,,$$
\item Cross coupling terms between the flat direction and inflaton superpotential:
$$- \frac{3}{M_p^2} W^*(s^\dagger) W(\phi) + h.c.\,,$$
\item Couplings between the flat direction and inflaton:
$$W_\phi K^{\phi\bar{s}} \left( D_{\bar{S}}W^*(s^\dagger) \right) + h.c\,.$$
\end{enumerate}
The resulting {\it leading order} potential for the light field $\phi$ at low energies can be captured by the following terms~\cite{Dine:1995kz,Enqvist:2003gh}:
%
%\begin{widetext}
\begin{equation}\label{reference}
 V(\phi)=V(s)+ (m_\phi^2+c_H H(t)^2)\lvert\phi\rvert^2+\bigg(A\frac{\lambda\phi^n}{nM_{p}^{n-3}}+a_H H(t) \frac{\lambda \phi^n}{nM_{p}^{n-3}}
 +h.c.\bigg)+\lambda^2\frac{\lvert\phi\rvert^{2(n-1)}}{M_{p}^{2(n-3)}} +\cdots \,,
\end{equation}
%\end{widetext}
%
where $A\sim m_\phi$ is the dimension full quantity, $\cdots$ contain terms of higher orders, $c_H,~a_H$ are numbers containing the information about the K\"ahler potential,
we can infer them from Table.~\ref{tab1}, and Appendix B and C. Note that during inflation, $H(t)\sim H_{inf}$, is nearly constant. Note that there are two kinds of 
Hubble-induced terms; one proportional to the mass term, and the second of the order of the $A$-term. 

\begin{table*}
\centering
\tiny\tiny
\begin{tabular}{|c|c|c|}
\hline
\hline
\hline
%Best fit Cosmological parameters
%\hline
% & &  &   \\
% & &  & \\
\footnotesize{\bf Non--minimal} & \footnotesize{\bf Non--canonical} & \footnotesize{\bf Potential}\\
\footnotesize{\bf K\"{a}hler potential} &\footnotesize{\bf kinetic term} & \footnotesize $V(\phi)$ \\
 & \footnotesize$\mathcal{L}_{Kin}=K_{ij^*}(\partial_\mu\Phi^{i})(\partial^\mu \Phi^{j^*})$ &  \footnotesize{\bf for } $ |s|<<M_{p}$   \\
%&  &  &  {\bf for } $ |I|<<M_{p}$\\
%& & & \\
\hline\hline\hline
%& & & \\ 
 \footnotesize $\displaystyle K^{(1)}= \phi^\dag \phi+s^\dag s$ & 
\footnotesize$\mathcal{L}_{Kin}=\displaystyle\bigg(1+\frac{a\lvert s\rvert^2}{M_{p}^2}\bigg)(\partial_\mu\phi)(\partial^\mu\phi^{\dagger})$
  &\footnotesize  $V(s)+\bigg(m_\phi^2+3(1-a)H^2\bigg)\lvert\phi\rvert^2\footnotesize \displaystyle-
 \frac{A\phi^n}{nM_{p}^{n-3}}$
\\
 \footnotesize $+\frac{a}{M_{p}^2}\phi^\dag \phi s^\dag s $ &\footnotesize $\displaystyle+\frac{a}{M_{p}^2}
\big\{\phi^{\dagger} s(\partial_\mu\phi)(\partial^\mu s^{\dagger})$  
&  \footnotesize $\displaystyle-\left( 1 + a \frac{|s|^2}{M_{p}^2} \right)\Big(1-\frac{3}{n}\Big)\frac{ s^2}{M^{2}_{p}}
\frac{\lambda M_{s}\phi^n}{M^{n-3}_{p}}$
\\
&\footnotesize $\displaystyle+\phi s^{\dagger}(\partial_\mu s)(\partial^\mu \phi^{\dagger})\big\}$ &  
\footnotesize$\displaystyle-\left( 1 - a \frac{|s|^2}{M_{p}^2} \right)\Big(a-\frac{1}{n}\Big) \frac{(s^\dagger)^2}{M^{2}_{p}} 
\frac{\lambda M_{s}\phi^n}{M^{n-3}_{p}}$
\\
 &\footnotesize $\displaystyle+\bigg(1+\frac{a\lvert \phi\rvert^2}{M_{p}^2}\bigg)(\partial_\mu s)(\partial^\mu s^{\dagger})$ & 
\footnotesize $\displaystyle+\lambda^2\frac{\lvert\phi\rvert^{2(n-1)}}{M_{p}^{2(n-3)}}+h.c.$\\

\hline
%& & & \\
 \footnotesize$\displaystyle K^{(2)}=\phi^\dag\phi +s^\dag s$
&  \footnotesize $\mathcal{L}_{Kin}=\displaystyle(\partial_\mu\phi)(\partial^\mu\phi^{\dagger})+(\partial_\mu s)(\partial^\mu s^{\dagger})$
 &   \footnotesize $V(s)+\bigg(m_\phi^2+3(1+b^2)H^2\bigg)\lvert\phi\rvert^2\displaystyle-A \frac{\phi^n}{nM_{p}^{n-3}}$
\\ 
\footnotesize $+\frac{b}{2M_{p}}s^\dag\phi \phi + h.c.$ & \footnotesize $\displaystyle+\frac{b \phi }{2M_{p}}(\partial_\mu\phi)(\partial^\mu s^{\dagger})$ 
&  \footnotesize $\displaystyle-\left\{\left(1-\frac{3}{n}\right)\phi +\frac{ b  \phi^\dagger  s}{n M_{p}} \right\}\frac{\lambda \phi^{n-1}M_{s} s^2}{M_{p}^{n-1}}$ 
\\ 
 &  \footnotesize $\displaystyle+\frac{b \phi^{\dagger}}{2M_{p}}(\partial_\mu s)(\partial^\mu\phi^{\dagger})$ 
& \footnotesize $\displaystyle-\Big(\frac{s^\dagger \phi}{M_{p}}-bn\phi^\dagger \Big)\frac{2M_{s}\lambda \phi^{n-1} s^{\dagger}}{n M^{n-2}_{p}}$  
\\
 & &  \footnotesize  $\displaystyle-\frac{b M_{s} s^2}{2M_{p}^2}  
\left( \frac{2M_{s} s}{M_{p}}-\frac{M_{s} s^2 s^\dagger}{M_{p}^3} \right) \phi \phi$
 \\
 & &  \footnotesize $\displaystyle-\frac{4M^{2}_{s} b |s|^2 s^\dagger}{M^3_{p}} \phi \phi\displaystyle+\lambda^2\frac{\lvert\phi\rvert^{2(n-1)}}{M_{p}^{2(n-3)}}+h.c.$\\
%& & & \\
\hline
%& & & \\
  \footnotesize $\displaystyle K^{(3)}=\phi \phi^\dag+ss^\dag$ & 
 \footnotesize $\mathcal{L}_{Kin}=\displaystyle(\partial_\mu\phi)(\partial^\mu\phi^{\dagger})+(\partial_\mu s)
(\partial^\mu s^{\dagger})$ & \footnotesize $V(s)+\bigg(m_\phi^2+3H^2\bigg)\lvert\phi\rvert^2\displaystyle-
 A\frac{\phi^n}{nM_{p}^{n-3}}$\\
  \footnotesize $+\frac{c}{4M_{p}^2}s^\dag s^\dag\phi \phi + h.c. $ &  \footnotesize $\displaystyle +\frac{c s^{\dagger}
 \phi}{4M_{p}^2}(\partial_\mu \phi)(\partial^\mu s^{\dagger})$ &  
  \footnotesize $\displaystyle-\left\{\left(1-\frac{3}{n}\right)\phi  +\frac{c\phi^\dagger  ss}{2M_{p}^2}\right\} \frac{\lambda \phi^{n-1}M_{s} s^2}{M_{p}^{n-1}}$ 
\\
 &  \footnotesize $\displaystyle+\frac{c s \phi^{\dagger}}{4M_{p}^2}(\partial_\mu s)(\partial^\mu\phi^{\dagger})$ &
 \footnotesize $\displaystyle+\frac{c M^{2}_{s} s^2s^\dagger s \phi\phi  }{M^4_{p}}-\frac{M^{2}_{s} c 
|s|^2 s^\dagger s^\dagger}{M^4_{p}} \phi \phi\displaystyle+\lambda^2\frac{\lvert\phi\rvert^{2(n-1)}}{M_{p}^{2(n-3)}}$
 \\
& &  \footnotesize $\displaystyle-\Big(\frac{s^\dagger \phi}{M_{p}}-\frac{cn\phi^\dagger s}{M_{p}} \Big)\frac{2M_{s}\lambda \phi^{n-1} s^{\dagger}}{n M^{n-2}_{p}}+h.c.$
\\
%& & \footnotesize $$\\
%& & & \\
\hline
%& & & \\
  \footnotesize$\displaystyle K^{(4)}=\phi \phi^\dag+ss^\dag$ 
& 
 \footnotesize $\mathcal{L}_{Kin}=\displaystyle\bigg(\frac{d s}{M_{p}}+\frac{d s^{\dagger}}{M_{p}}+1\bigg)(\partial_\mu\phi)(\partial^\mu\phi^{\dagger})$ 
&  \footnotesize $V(s)+\bigg(m_\phi^2+3(1+d^2)H^2\bigg)\lvert\phi\rvert^2\displaystyle-A\frac{\phi^n}{nM_{p}^{n-3}}$\\
   \footnotesize $+\frac{d}{M_{p}}s\phi^\dag \phi + h.c.$ &  \footnotesize $\displaystyle+(\partial_\mu s)(\partial^\mu s^{\dagger})$ 
 &  \footnotesize $\displaystyle-\left(1-\frac{3}{n}\right) \frac{\lambda \phi^{n}M_{s} s^2}{M_{p}^{n-1}}
\displaystyle-\Big(\frac{s^\dagger}{M_{p}} -d\Big)\frac{2M_{s}\lambda \phi^{n} s^{\dagger}}{n M^{n-2}_{p}}$
 \\
 &  \footnotesize $\displaystyle+\frac{d \phi^{\dagger}}{M_{p}}(\partial_\mu \phi)(\partial^\mu s^{\dagger})
\displaystyle+\frac{d \phi }{M_{p}}(\partial_\mu s)(\partial^\mu\phi^{\dagger})$ 
&  \footnotesize $\displaystyle+\lambda^2\frac{\lvert\phi\rvert^{2(n-1)}}{M_{p}^{2(n-3)}}+h.c.$
\\
%& & & \\
\hline
\hline
\hline
\end{tabular}
\vspace{.4cm}
\caption{\label{tab1}
 Various supergravity effective potentials and non-canonical kinetic terms for $|s|<<M_{p}$ in presence non-niminmal K\"{a}hler potential. Here both $\phi$ and $s$ are complex fields, and so are the $A$-terms.} 
\end{table*}

%%%%%%%%%%%%%%%%%%%%%%%%%%%%%%%%%%%%%%%%%%%%%%%%%%%%%%%%%%%%

\section{Non--minimal K\"{a}hler potential and non--canonical kinetic terms}\label{N-M}

In this section we will consider two interesting possibilities, one which is the simplest and provides an excellent model 
for inflation with a complete decoupling of the heavy field. Inflation occurs via the slow roll of $\phi$ field within an MSSM vacuum, where inflation would end in a vacuum 
with an  {\it enhanced gauge symmetry}, where the entire electroweak symmetry will be completely restored.

\subsection{Heavy field is {\it dynamically} frozen}\label{frozen}

Let us first assume that the dynamics of the heavy field $s$ is completely frozen during the onset and the rest of the course of slow roll inflation driven by $\phi$. 
The full potential can be found in Table~\ref{tab1}. Note that
the potential for $s$ field, $V(s)$ contains  soft term and the corresponding $A$-term:
\begin{equation}
V(s)\sim M_s^2|s|^2+A'M_ss^2\,,
\end {equation}
where $A'$ is a dimensional quantity, and it is roughly proportional to $A'\sim M_s\gg {\rm TeV}$.
In this case there are two possibilities which we briefly mention below:

\begin{itemize}

\item We can imagine that the heavy field, $s$,  to have a  
global minimum at:
\begin{equation}
\langle s \rangle =0,~~~~~~~~~~~\langle \dot s \rangle =0\,,~~~~~~~~V(s)=0\,.
\end{equation}
In this particular setup, the kinetic terms for each cases, i.e. $1, 2, 3, 4$, become canonical for the $\phi$ field,  therefore the heavy field is completely decoupled from the dynamics. 
One can check them from Table-\ref{tab1}.
This is most ideal situation for a single field dominated model of inflation, where the overall potential for 
along $\phi$ direction simplifies to:
\begin{equation}
V(\phi) =m_\phi^2\lvert\phi\rvert^2+\bigg(A\frac{\lambda\phi^n}{nM_{p}^{n-3}}
 +h.c.\bigg)+\lambda^2\frac{\lvert\phi\rvert^{2(n-1)}}{M_{p}^{2(n-3)}}\,.
\end{equation}
The overall potential is solely dominated by the $\phi$ field, therefore Hubble expansion rate becomes, 
$H_{inf} \propto V(\phi)/M_{p}^2$. 

In this setup inflation can occur near a saddle point or an inflection point, where $\phi_0\ll M_p$, and $m_\phi \gg H_{inf}$, first discussed in Refs.
~\cite{Allahverdi:2006we,Allahverdi:2006iq}. During inflation the Hubble 
expansion rate is smaller than the soft SUSY breaking mass term and the $A$-term, i.e. $A\sim m_\phi \gg H(t)$ for $a_H \sim c_H\sim {\cal O}(1)$ in Eq.~(\ref{reference}), such that the SUGRA corrections are unimportant. This scenario has been discussed extensively, and has been extremely successful with the Planck data explaining the spectral tilt right on the 
observed central value, with Gaussian perturbations with the right amplitude~\cite{Wang:2013hva,Choudhury:2013jya}. 

\item On the other hand, if 
\begin{equation}
\langle s\rangle\sim M_s \ll M_p\,,~~~~~~~~~~~ \langle \dot s\rangle =0\,,~~~~~~~~~~~V(s)=M_s^4\,,
\end{equation}
then the kinetic term for $\phi$ field will be canonical for cases $K^{2}$ and $K^{3}$ by virtue $\dot s =0$, see Table-\ref{tab1}.
However for cases $K^{1}$ and $K^{4}$, the departure from canonical for the $\phi$ field will depend on $M_s$. If
$\langle M_s\rangle \ll M_p$,  and $ a,~d \sim {\cal O}(1)$, see Table-\ref{tab1}, then the kinetic term for $\phi$ will be virtually
canonical, and as a consequence $c_s\approx 1$, while the potential will see a modification:
\begin{equation}
V_{total} = M_s^4+c_{H}H^{2}|\phi|^{2}+ \bigg(a_H H\frac{\lambda\phi^n}{nM_{p}^{n-3}}
 +h.c.\bigg)+\lambda^2\frac{\lvert\phi\rvert^{2(n-1)}}{M_{p}^{2(n-3)}}\,.
\end{equation}
This large vacuum energy density, i.e. $M_s^4\gg (\rm TeV)^4$, would  yield a large Hubble expansion rate, i.e. $H_{inf}^2\sim M_s^4/M_p^2\gg m^2_\phi\sim {\cal O}({\rm TeV})^2$. 
Therefore, the Hubble induced mass and and the $A$-term would dominate the potential over the soft terms.  Inspite of large mass, $c_H$, and  $a_{H}$-term, there is {\it no} SUGRA-$\eta$
problem, provided inflation occurs near the {\it saddle point} or the {\it inflection point}~\cite{Mazumdar:2011ih,Choudhury:2013jya}. We will not discuss this case any further, we will
now focus on a slightly non-trivial scenario, where high scale physics can alter some of the key cosmological predictions. 

\end{itemize}

 %%%%%%%%%%%%%%%%%%%%%%%%%%%%%%%%%%%%%%%%%%%%%%%%

 \subsection{Heavy field is oscillating during the onset of inflation}\label{oscillating}

One dramatic way the heavy field can influence the dynamics of primordial perturbations is via
coherent oscillations around its minimum,  while $\phi$ still plays the role of a slow roll inflaton~\footnote{There could
be other scenarios where the influence of heavy field is felt throughout the inflationary dynamics, see for instance in 
Refs.~\cite{Achucarro:2010da,Achucarro:2010jv,Assassi:2013gxa,Rehman:2010wm,Shafi:2010jr,Khalil:2010cp,Antusch:2010mv,Antusch:2012jc}. Here 
we will discuss a slightly simpler scenario where both heavy and light sectors are coupled gravitationally via the K\"ahler correction.}. Furthermore,  the heavy field would only influence the first few e-foldings of inflation, once the heavy field 
is settled down its effect would be felt only via the vacuum energy density. Inspite of this short-lived phase, the heavy field can influence the dynamics and the perturbations for
the light field as we shall discuss below. 

Let us imagine the heavy field is coherently oscillating around a VEV, $\langle s\rangle \sim M_s$, during the 
initial phase of inflation, such that
\begin{equation}
V(s)\neq 0,~~~~~~~~~~~~~~~~\langle s\rangle \neq 0,~~~~~~~~~~~~~~~~\langle \dot  s\rangle \neq 0\,.
\end{equation}
The origin of coherent oscillations of $s$ field need not be completely ad-hoc, such a scenario might arise quite naturally from the hidden sector moduli field which is  coherently oscillating before being damped away by the initial phase of inflation, see for instance~\cite{Douglas:2006es}. This is particularly plausible for high string scale moduli, where the moduli mass can be heavy and can be stabilised early on in the history of the universe. There could also be a possibility of a smooth second order phase transition from one vacuum to another
during the intermittent phases of inflation~\cite{Burgess:2005sb,Allahverdi:2007wh}. Such a possibility can arise within MSSM where there are multiple false vacua  at high energies~\cite{Allahverdi:2008bt}. Irrespective of the origin of this heavy field,
during this transient period, the heavy field with an effective mass, $M_s \gg H_{inf}$, can coherently oscillate around its vacuum. We can set its 
initial amplitude of the oscillations to be of the order $M_s$. 
\begin{equation}\label{heavyf}
 s(t)=M_{s}+M_{s}\sin(M_st)\,.
 \end{equation}
This also implies that at the lowest order approximation, $\langle s \rangle \sim M_{s}$ and $\langle \dot s\rangle\sim M^{2}_{s}$~\footnote{At this point one might
say why we had taken the amplitude of oscillations for the heavy field to be $M_s$. In some scenarios, it is possible to envisage the amplitude of the oscillations to be $M_p$.
This would not alter much of our discussion, therefore for the sake of simplicity we will consider the initial amplitude for the $s$ field to be displaced by $M_s$, the same as that of the VEV.}.
The contribution to the potential due to the time dependent oscillating heavy field, see Eq~(\ref{heavyf}), is averaged over a full cycle ($0<t_{osc}<H_{inf}^{-1}$) is given by:
%\begin{widetext}
\be\begin{array}{lll}\label{avg}
    \displaystyle \langle V(s) \rangle \approx M_s^{2} \langle s^{2}(t) \rangle \sim H_{inf}^{2} M^{2}_{p}
   \end{array}
\ee
%\end{widetext}
The $s$ field provides at the lowest order corrections to the kinetic term for the $\phi$ field, 
and to the overall potential, see Table~(\ref{tab1}), for both kinetic and potential  terms. 

At this point one might worry, the coherent oscillations of the $s$ field might trigger particle creation from 
the time dependent vacuum, see Refs.~\cite{Traschen:1990sw,Kofman:2004yc}, for a review see~\cite{Allahverdi:2010xz}. 
First of all, if we assume that the heavy field is coupled to other fields gravitationally, then the particle creation may not be 
sufficient to back react into the inflationary potential. Furthermore, inflation would also dilute the quanta created during this 
transient phase. We would not expect any imprint of this event on cosmological scales~\cite{Biswas:2013lna}, except one interesting possibility could be 
to excite some non-Gaussianity~\cite{Enqvist:2004ey,Biswas:2013fna}. In this paper we will not study the effects of non-Gaussianity, we shall leave this 
question for the companion paper.

Since the kinetic terms for the $4$ cases tabulated in Table~(\ref{tab1}) are now no longer canonical, they would contribute to 
the speed of sound, $c_s \neq 1$, which we can summarize case by case below:
%
%\begin{widetext}
\be\label{sigbz}
 c_s=\sqrt{\frac{\dot p}{\dot \rho}}\approx \left\{
	\begin{array}{ll}
                    \displaystyle \sqrt{\frac{{\bf X}_{1}(t)-{\bf X}_{2}(t)-\dot{\widehat{V}}}{{\bf X}_{1}(t)+{\bf X}_{3}(t)+\dot{\widehat{V}}}} & \mbox{ for $\underline{\bf Case ~I}$}  
                    \\\\
         \displaystyle \sqrt{\frac{{\bf Y}_{1}(t)-{\bf Y}_{2}(t)-\dot{\widehat{V}}}{{\bf Y}_{1}(t)+{\bf Y}_{3}(t)+\dot{\widehat{V}}}} & \mbox{ for $\underline{\bf Case ~II}$}\\\\
\displaystyle \sqrt{\frac{{\bf Z}_{1}(t)-{\bf Z}_{2}(t)-\dot{\widehat{V}}}{{\bf Z}_{1}(t)+{\bf Z}_{3}(t)+\dot{\widehat{V}}}} & \mbox{ for $\underline{\bf Case ~III}$}\\\\
\displaystyle \sqrt{\frac{{\bf W}_{1}(t)-{\bf W}_{2}(t)-\dot{\widehat{V}}}{{\bf W}_{1}(t)+{\bf W}_{3}(t)+\dot{\widehat{V}}}} & \mbox{ for $\underline{\bf Case ~IV}$} .
          \end{array}
\right. 
\ee
%\end{widetext}
%
where $p$ is the effective pressure and $\rho$ is the energy density. The dot denotes derivative w.r.t. physical time, $t$.
All the symbols, i.e. $X_1,~X_2,~Y_1,~Y_2,~Z_1,~Z_2,~W_1,~W_2$, appearing in Eq~(\ref{sigbz}) are explicitly mentioned in 
the appendix. Additionally, here we have defined, $\widehat{V}=V(\phi)-V(s)$~\footnote{ As a side remark, our analysis will be very useful
for the Affleck-Dine (AD) baryogenesis~\cite{Dine:1995kz}, especially when the minimum of the AD field is rotating in presence of the inflaton 
oscillations. Effectively, the AD field will have non-canonical kinetic terms, this has never been taken into account in the literature and one
should take the non-canonical kinetic terms for the AD field in presence of the inflaton oscillations in order to correctly estimate the baryon asymmetry.
The role of $s$ field will be that of an inflaton and $\phi$ field will be that of an AD field.}.
%%%%%%%%%%%%%%%%%%%%%%%%%%%%%%%%%%%%%%%%%%%%

\begin{figure*}[htb]
\centering
\subfigure[$\textcolor{blue}{\bf \underline{Case ~I}}$]{
    \includegraphics[width=7.2cm,height=6.5cm] {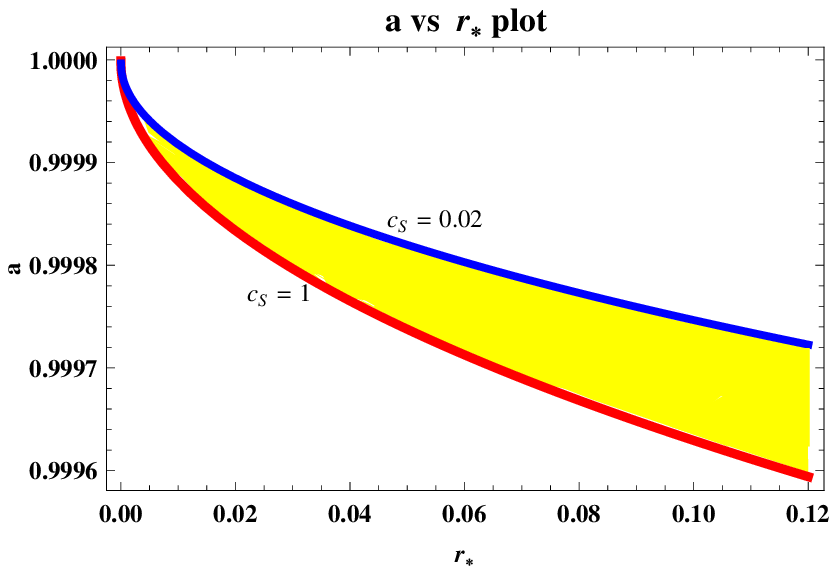}
    \label{fig:subfig1}
}
\subfigure[$\textcolor{blue}{\bf \underline{Case ~II}}$]{
    \includegraphics[width=7.2cm,height=6.5cm] {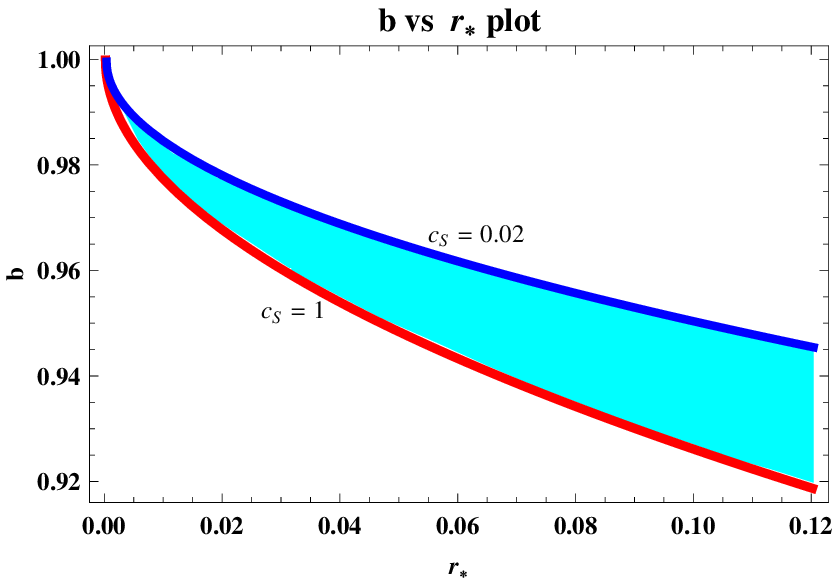}
    \label{fig:subfig2}
}
\subfigure[$\textcolor{blue}{\bf \underline{Case ~III}}$]{
    \includegraphics[width=7.2cm,height=6.5cm] {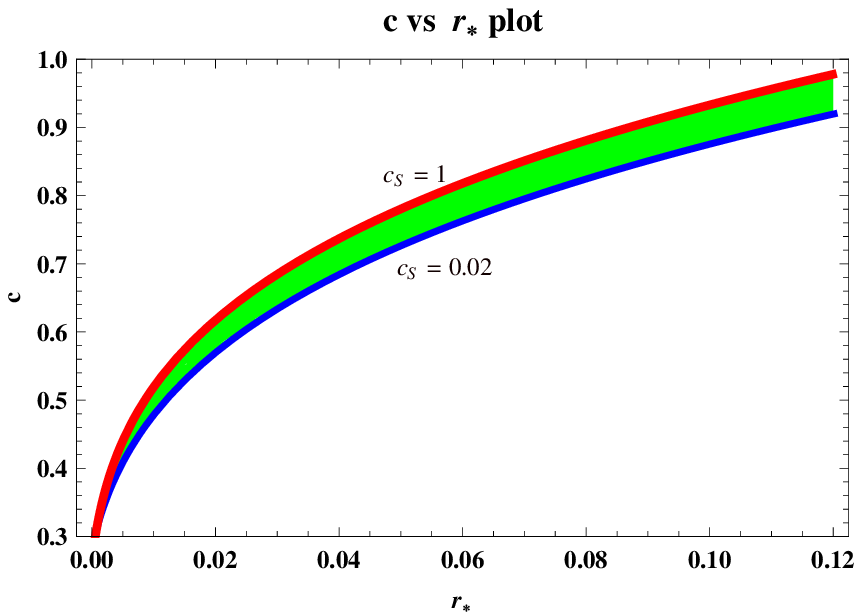}
    \label{fig:subfig3}
}
\subfigure[$\textcolor{blue}{\bf \underline{Case ~IV}}$]{
    \includegraphics[width=7.2cm,height=6.5cm] {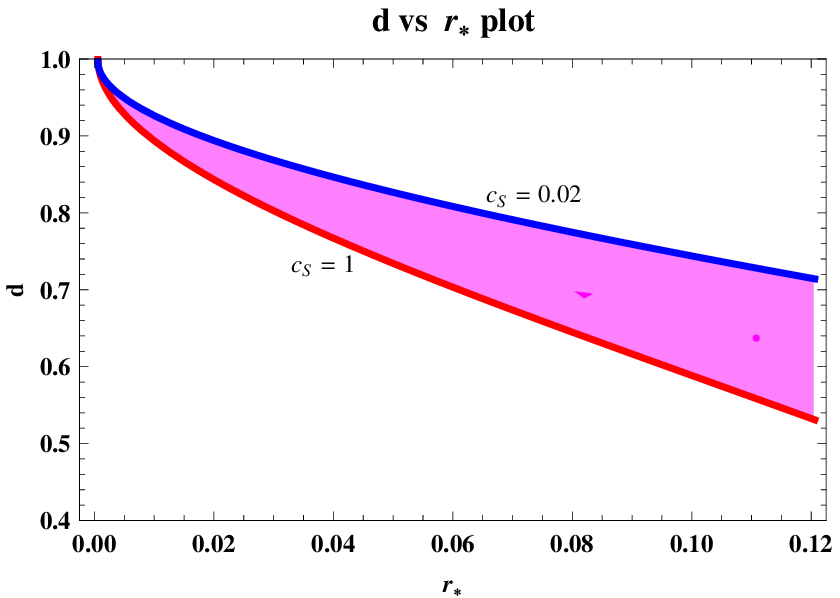}
    \label{fig:subfig4}
}
\caption[Optional caption for list of figures]{We show the constraints on 
  the non-renormalizable  K\"ahler operators, ``a'',``b'',``c'' and ``d'' 
with respect to the tensor-to-scalar ratio $r_{\star}$ at the pivot scale $k_{\star}=0.002~Mpc^{-1}$
when the heavy field $s$ is oscillating during the initial phase of inflation, especially at the time 
when the interesting perturbations are leaving the Hubble patch for $H_{inf} >>m_{\phi}\sim{\cal O}({\rm TeV})$. 
All the shaded regions represent the allowed parameter space for the Hubble induced inflation satisfying
the Planck $2\sigma$ constraints on the amplitude of power spectrum $2.092\times 10^{-9}<P_S< 2.297\times 10^{-9}$ and spectral tilt $0.958 < n_S<0.963$,
as mentioned in Eq~(\ref{powq1}) and Eq~(\ref{powq2}) respectively.
 The dark coloured boundaries are obtained from the allowed range of the 
speed of sound $c_{s}$, within the window $0.02\leq c_{s}\leq 1$. }  
\label{fz}
\end{figure*}

%%%%%%%%%%%%%%%%%%%%%%%%%%%%%%%

\begin{figure*}[htb]
\centering
\subfigure[$P_{S}$~vs~$n_{S}$]{
    \includegraphics[width=7.3cm,height=6.0cm] {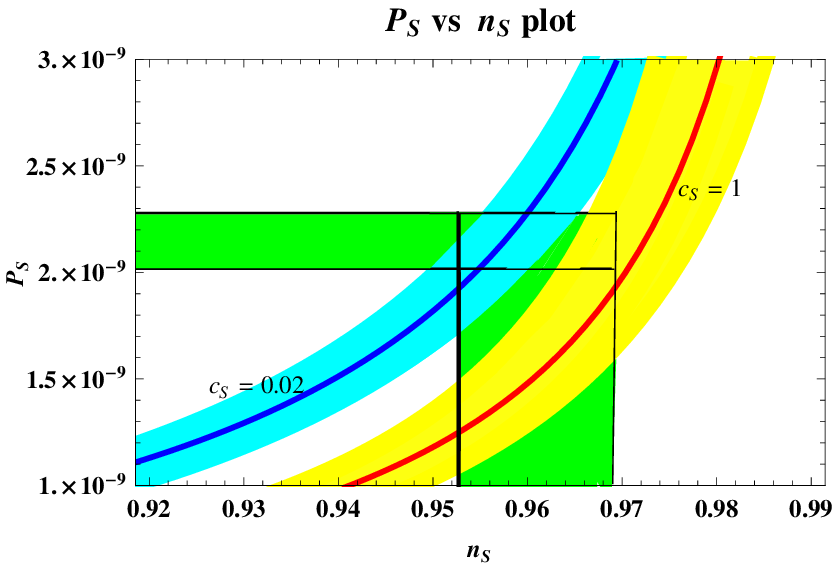}
    \label{fig:subfig5}
}
\subfigure[$\log(r)$~vs~$n_{S}$]{
    \includegraphics[width=7.3cm,height=6.0cm] {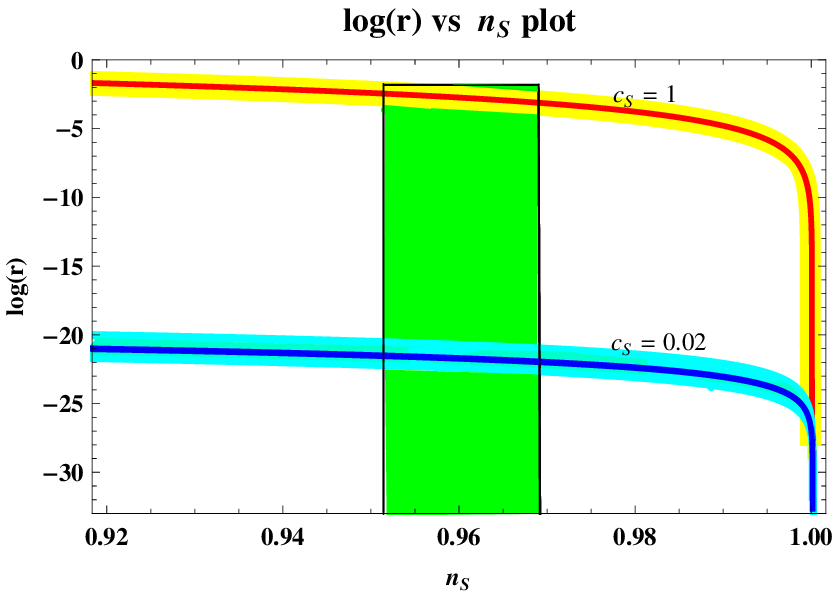}
    \label{fig:subfig6}
}
\caption[Optional caption for list of figures]{For $H_{inf}\geq m_\phi\sim {\cal O}({\rm TeV})$, we have shown the variation of 
\subref{fig:subfig5} $P_{S}$~vs~$n_{S}$, and \subref{fig:subfig6} $\log(r)$~vs~$n_{S}$, for MSSM flat direction (for $\widetilde u\widetilde d\widetilde d$ and $\widetilde L\widetilde L\widetilde e$ inflaton cadidates)
 in presence of non-minimal K\"ahler corrections. The {\it red} and {\it blue} curves are
drawn for $c_{s}=1$ and $c_{s}=0.02$, which show the
model parameters, $\delta\sim 10^{-4},~\lambda =1,~c_H=2,~a_H=2.108,~\phi_0=1.129\times 10^{16}~{\rm GeV}$, for the pivot scale 
$k_{\star}=0.002~Mpc^{-1}$. The ${\it green}$ shaded region shows the  $2\sigma$ CL. range 
allowed by the Planck data~\cite{Planck-infl} for both the cases. Instead of getting a solid {\it red} and {\it blue} curves we obtain a {\it yellow} and {\it aqua} 
shaded regions if we would consider the full parameter space for ($H_{inf}\geq m_{\phi}\sim {\cal O}({\rm TeV})$).
}
\label{fgh}
\end{figure*}

%%%%%%%%%%%%%%%%%%%%%%%%%%%%%%%%%%%%%%%%%%%%%%%%%%%%%%%%%%%%

%%%%%%%%%%%%%%%%%%%%%%%%%%%%%%%%%%%%%%%%%%%%%%%%

 \subsection{Constraining non-renormalizable operators, i.e. $a,~b,~c,~d$, and $M_s$}\label{02}
 
For the potential under consideration, we have $V(s)=3H^2M^{2}_{p}\sim M^{2}_{s}s^2>>m_{\phi}^2|\phi|^2$, 
where $m_{\phi}\sim{\cal O}(\rm TeV)$ is the soft mass. In this case the 
contributions from the Hubble-induced terms are important compared to the soft SUSY breaking mass, $m_\phi$, and the $A$ term for 
all the four cases tabulated in Table-(\ref{tab1}).
The potential, Eq.~(\ref{reference}), after stabilizing the angular direction of the complex scalar field $\phi= |\phi|\exp[i\theta]$, see~\cite{Allahverdi:2006we,Allahverdi:2006iq,Allahverdi:2006cx,Mazumdar:2011ih}, reduces to a simple form along the real direction, which is dominated by a 
single scale, i.e. $H\sim H_{inf}$:
\begin{equation}\label{h1a}
 V(\phi)=V(s)+c_{H}H^{2}|\phi|^{2}-
 \frac{a_{H}H\phi^n}{nM_{p}^{n-3}}+\frac{\lambda^2\lvert\phi\rvert^{2(n-1)}}{M_{p}^{2(n-3)}},
\end{equation}
where we take $\lambda=1$, and,the Hubble-induced mass parameter $c_{H}$, for $s<<M_{p}$, is defined as~\footnote{See Appendix-B and Appendix-C for details.}:
\be\label{effective}
c_{H}=\left\{
	\begin{array}{ll}
                    \displaystyle  3(1-a)\,, & \mbox{ for $\underline{\bf Case ~I}$}  \\
         \displaystyle  3(1+b^2)\,, & \mbox{ for $\underline{\bf Case ~II}$}\\
\displaystyle  3\,, & \mbox{ for $\underline{\bf Case ~III}$}\\
\displaystyle  3(1+d^2)\,, & \mbox{ for $\underline{\bf Case ~IV}$}.
          \end{array}
\right.
\ee
Note that for only third case, i.e. $K^{3}$, the Hubble induced mass term does not contain any K\"ahler correction, i.e. $\delta K$.
Similarly, we can express $a_H$, see Appendix~C for full expressions.
Note that for all $4$ cases, the kinetic terms are all non-minimal, and we have already listed in Table-(\ref{tab1}).
Fortunately for this class of potential given by Eq~(\ref{h1a}), inflection point inflation can be accommodated, when  $a_H^2\approx 8(n-1)c_H$. 
This can be characterized by a fine-tuning parameter, $\delta$, which is defined as~\cite{Allahverdi:2006we}:
\begin{equation}
\label{newbeta}
\frac{a_H^2}{8(n-1)c_H} = 1-\left(\frac{n-2}{2}\right)^2\delta^2\,.
\end{equation}
When $\vert\delta\vert$ is small~\footnote{We will consider a moderate tuning of order $\delta \sim 10^{-4}$ between $c_H$ and $a_H$.}, a point of inflection $\phi_0$ exists,
such that $V^{\prime\prime}\left(\phi_0\right) =0$, with
\begin{equation}
\label{phi0}
\phi_0 = \left(\sqrt{\frac{c_H}{(n-1)}} H M_{p}^{n-3}\right)^{{1}/{n-2}}\, +{\cal O}(\delta^2).
\end{equation}

For $\delta <1$, we can Taylor-expand the inflaton potential around an inflection 
point, $\phi=\phi_{0}$, as~\cite{Enqvist:2010vd,Mazumdar:2011ih,Hotchkiss:2011am}:
\be\label{rt1a}
V(\phi)=\alpha+\beta(\phi-\phi_{0})+\gamma(\phi-\phi_{0})^{3}+\kappa(\phi-\phi_{0})^{4}+\cdots\,,
\ee 
where the expansion coefficients are now given by:
%\begin{widetext}
\begin{eqnarray}\label{p1}
     \alpha&=&V(\phi_{0})=V(s)+\left(\frac{(n-2)^{2}}{n(n-1)}+\frac{(n-2)^2}{n}\delta^{2}\right)c_{H}H^{2}\phi^{2}_{0}+{\cal O}(\delta^{4}),\\
 \beta&=&V^{'}(\phi_{0})=2\left(\frac{n-2}{2}\right)^{2}\delta^{2}c_{H}H^{2}\phi_{0}+{\cal O}(\delta^{4}),\\
 \gamma&=&\frac{V^{'''}(\phi_{0})}{3!}=\frac{c_{H}H^{2}}{\phi_{0}}\left(4(n-2)^2-\frac{(n-1)(n-2)^3}{2}\delta^{2}\right)+{\cal O}(\delta^{4}),\\
 \small\kappa&=&\frac{V^{''''}(\phi_{0})}{4!}\\~~~~&=&\frac{c_{H}H^{2}}{\phi^{2}_{0}}\left(12(n-2)^3-\frac{(n-1)(n-2)(n-3)
(7n^2-27n+26)}{2}\delta^{2}\right)+{\cal O}(\delta^{4}).\nonumber 
   \end{eqnarray}
%\end{widetext}
%
Note that once we specify $c_H$ and $H_{inf}$, all the terms in the potential are determined. In this regard the potential 
indeed simplifies a lot to study the cosmological observables.

As an concrete example, we considered $n=6$ case, where the flatness of the superfield $\Phi$ is lifted by the non-renormalizable operator.
This is appropriate for both $\widetilde u\widetilde d\widetilde d$ and $\widetilde L\widetilde L\widetilde e$ flat directions.
In our scans we allow the constraints from Planck observations~\cite{Planck-infl,Planck-1}, see Eqs.~(\ref{powq1},\ref{powq2},\ref{c-s},\ref{t-s}).

Let us now scan the parameter space for $c_H,~a_H$ with the help of
Eqs.~(\ref{scalar},~\ref{tensor},~\ref{scalartilt},~\ref{tensortilt},~\ref{sconfo}),  by fixing $\lambda ={\cal O}(1)$ and $\delta \sim 10^{-4}$. 
In order to satisfy the Planck observational constraints on the amplitude of the power spectrum, $2.092\times 10^{-9}<P_S< 2.297\times 10^{-9}$ ( within $2\sigma$),
spectral tilt $0.958 < n_S<0.963$ ( within $2\sigma$ ), sound speed $0.02\leq c_s\leq 1$ ( within $2\sigma$), and tensor-to-scalar ratio $r_\star \leq 0.12$, we obtain the following
constraints on our parameters for $H_{inf}\geq m_{\phi}\sim {\cal O}(\rm TeV)$, where successful inflation can occur via {\it inflection point} inflation:
\begin{eqnarray}\label{P-space}
%\delta& \sim &{\cal O}(10^{-4})\,,\nonumber \\
c_{H} &\sim &{\cal O}(10-10^{-6})\,, ~~~~~~~~~~~~~~~{\bf for}~~~~~~~~~~10^{-22}<r_{\star}<0.12\nonumber \\
a_{H} & \sim &{\cal O}(30 - 10^{-3} )\,, ~~~~~~~~~~~~~~~{\bf for}~~~~~~~~~~10^{-22}<r_{\star}<0.12\nonumber \\
M_s &\sim & {\cal O}(9.50\times 10^{10}-1.77\times10^{16})~{\rm GeV}\,,~~~{\bf for}~~~~10^{-22}<r_{\star}<0.12\,.
\end{eqnarray}
%\end{widetext}
Our motivation of doing such a scan is to generate feasible amplitude of power spectrum $P_s$ , spectral tilt $n_s$, sound speed $c_s$ 
and tensor to scalar ratio $r_\star$, which also satisfies the particle physics constraints in our prescribed inflationary setup. As these constraints are necessary
 to satisfy the inflation, we have to choose the parameter space in such a way that all of these constraints satisfy simultaneously. 
Inflation would not occur outside our scanning region since at least one of the constraints would be violated.

Note that for the above ranges, Eq.~(\ref{P-space}), $\phi_0$ gets automatically fixed by Eq.~(\ref{phi0}), 
\begin{equation}\label{pj0}
\phi_{0} \sim {\cal O}(10^{14} -10^{17})~{\rm GeV}\,~~~{\bf for}~~~~10^{-22}<r_{\star}<0.12\,.
\end{equation}
Here the upper and lower bound appearing in Eq~(\ref{P-space}) and Eq~(\ref{pj0})
are obtained from large and small values of the tensor-to-scalar ratio varying
within a wide range $10^{-22}<r_{\star}<0.12$ for the pivot scale $k_{\star}\sim 0.002~Mpc^{-1}$.

In Fig.~(\ref{fz}), we have shown that the allowed ranges of the non-renormalizable coefficients of
 the operators mentioned in Eqs.~(\ref{corrt-1},~\ref{corrt-2},~\ref{corrt-3},~\ref{corrt-4}).
 The solid blue and red 
curves are drawn for the sound speed $c_{s}=0.02$ and $c_{s}=1$ and the shaded regions are shown to point 
out the allowed region which satisfies the Planck $2\sigma$ constraints on the amplitude of power spectrum $P_S$ and spectral tilt $n_S$
as mentioned in Eq~(\ref{powq1}) and Eq~(\ref{powq2}) respectively. It is important to 
note that the non-minimal couplings ``a'', ``b'' and ``d'' directly controls both $c_H,~a_H$ in the inflaton potential. But the coupling ``c'' only 
affect $a_H$, while leaving $c_H$ free from non-minimal correction, i.e. $c_{H}\sim 3$. For the consistency check see appendix where all the non-minimal
couplings ``a'', ``b'', ``c'' and ``d'' are explicitly written in terms of the scale (VEV) of the heavy field $M_{s}$.

In Fig.~(\ref{fig:subfig5}) and Fig.~(\ref{fig:subfig6}) we have shown the constraints on the amplitude of the the power spectrum for scalar modes, $P_{S}$, and 
 $\log(r)$, with respect to spectral tilt, $n_{S}$
 at the pivot scale $k_{\star}=0.002~{\rm Mpc}^{-1}$ by {\it red} and {\it blue} curves for the sound speed, $c_{s}=1$ and $c_{s}=0.02$, respectively.
If we consider the full parameter space as stated in Eq.~(\ref{P-space}), there are solutions which have been shown in a {\it yellow} and {\it aqua} 
shaded regions. We have also shown the $2\sigma$ region 
allowed by the Planck data~\cite{Planck-infl} for both the cases by ${\it green}$ shaded region, i.e. $P_S$ and $~n_S$.
It is important to note that if we consider the full parameter space then the low $c_s$ fits the data well compared to the high value $c_s$.

\begin{figure}[t]
{\centerline{\includegraphics[width=8.5cm, height=6.2cm] {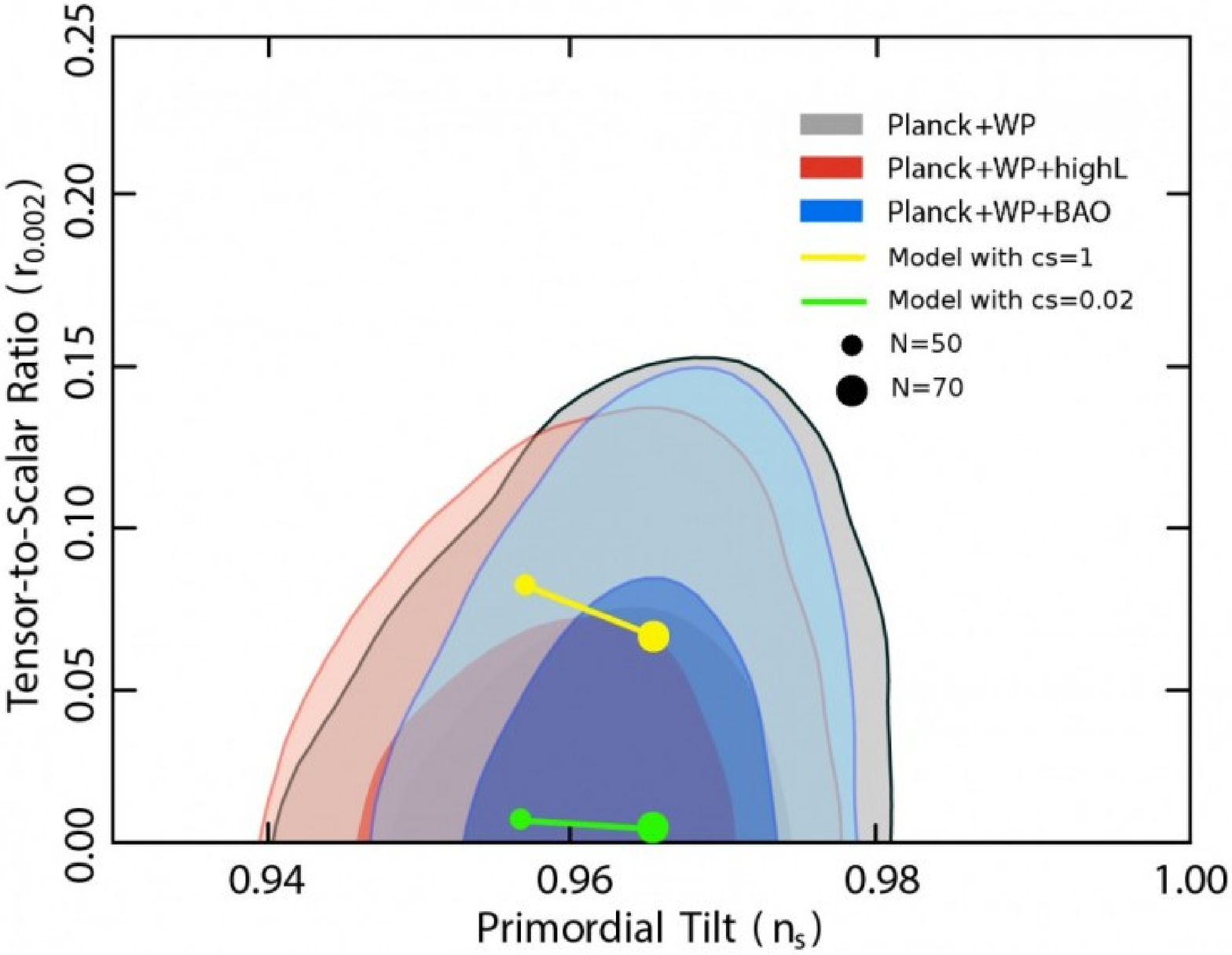}}}
\caption{We show the joint $1\sigma$ and $2\sigma$ CL. contours using
 {\it Planck}+WMAP-9, {\it Planck}+WMAP-9+high~{\it l} and {\it Planck}+WMAP-9+BAO data.
The {\it yellow} and {\it green } lines are drawn for the proposed model with $c_{s}=1$ and $c_{s}=0.02$ where the
model parameters are fixed at, $\delta\sim 10^{-4},~\lambda =1,~c_H=2,~a_H=2.108,~\phi_0=1.129\times 10^{16}~{\rm GeV}$, for the pivot scale 
$k_{\star}=0.002~Mpc^{-1}$ respectively. The 
 region in between the {\it yellow} and {\it green} lines represent the allowed region obtained from the model. 
The small circle on the left corresponds to $N=50$, while the right big circle corresponds to $N=70$.} \label{fig2con}
\end{figure}

However from Fig~(\ref{fig:subfig5}) it is clearly observed that the high value of $c_s$ also confronts the data well within a small patch for a specific choice of parameter
space lying within the parameter scanning range mentioned in Eq~(\ref{P-space}). Consequently the full parameter space for low $c_s$ and a tiny patch for high $c_s$ 
fits the CMB power spectra well in the low~${\it l}$ $(2<l<49)$ and high~${\it l}$ $(50<l<2500)$ multipole region. But for the low~${\it l}$ $(2<l<49)$
region, the statistical error is too huge to differentiate between different $c_s$ scenarios. Therefore, we will concentrate only on
 the high~{\it l} $(50<l<2500)$ region for the low and high $c_s$ model discrimination with high statistical accuracy ($2\sigma$ C.L.). See Fig~(\ref{fig2con})
for the details where we explicitly use this additional input. 

 %%%%%%%%%%%%%%%%%%%%%%FIGURE%%%%%%%%%%%%%%%%%%%%%%%%%%%%%%%%%%%%%

%%%%%%%%%%%%%%%%%%%%%%%%%%%%%%%%%%%%%%%%%%%%%%%%%%%%%%%%%%%%%%%%%%%%%%%%%%%%%%%%%%

Furthermore, by using {\it Planck}+WMAP-9~\cite{Planck-infl,Planck-1}, {\it Planck}+WMAP-9+high~{\it l} \cite{Planck-infl,Planck-1} and {\it Planck}+WMAP-9+BAO datasets~\cite{Planck-infl,Planck-1},
we have shown $r$ vs. $n_s$ in the marginalized $1\sigma$ and $2\sigma$ CL. contours  in Fig.~(\ref{fig2con}). The {\it yellow} and {\it green } lines are drawn for the proposed model with $c_{s}=1$ and $c_{s}=0.02$ respectively.
 The region in between the {\it yellow} and {\it green} lines represent the allowed region obtained from the proposed model within the window $50\leq N\leq70$.
In Fig~(\ref{fig2con}) we fix the number of e-foldings within the window, $50\leq N\leq70$, because at $N=50$ and $N=70$ the illustrated model satisfies the
Planck $2\sigma$ combined constraints on the upper and lower bound of the amplitude of the power spectrum $P_S$, spectral tilt $n_S$, and the upper bound of 
tensor-to-scalar ratio $r_\star$ as mentioned in Eq~(\ref{powq1},\ref{powq2},\ref{t-s}) at the pivot scale $k_{\star}\sim 0.002~{\rm Mpc}^{-1}$ for both $c_s=0.02$ and $c_s=1$ branch.

Let us now derive an analytical expression for the scale of inflation, i.e.. $M_s$. We consider a full cycle averaged within an interval 
$0<t_{osc}<H^{-1}_{inf}$, and using Eq~(\ref{phi0}), Eq~(\ref{p1}) and Eq~(\ref{hubble}) for n=6 flat directions, 
we can derive a following constraint on the scale of the heavy field, $M_s$ for $k_{\star}(\sim 0.002~Mpc^{-1})$, by
setting $\alpha \sim V(s) \approx M^4_{s}$ and the fine tuning parameter, $\delta \sim {\cal O}(10^{-4}) \ll 1$, the leading order contribution to the potential will be given by:
%
%\begin{widetext}
 \be\begin{array}{llll}\label{hscalecon1}
     \displaystyle M_{s}\leq 1.77\times 10^{16}\left({\frac{r_{\star}}{0.12}}\right)^{1/4}c^{\frac{\epsilon_{V}}{2(\epsilon_{V}-1)}}_{s}~{\rm GeV}\,.
%{\bf G_{\star}}
    \end{array}
\ee
The above Eq~(\ref{hscalecon1}) will fix the upper bound on $M_s\sim {\cal O}(10^{16})$ by setting $c_s=1$~\footnote{
In the setup $\epsilon_{V}(k_{*}\approx k_{cmb})\approx 0.0021$ which satisfies the WMAP+Planck constrain, as this combined data set puts an upper bound at $\epsilon_{V}<0.008 $at $95\%$ CL.~\cite{Planck-1,Ade:2013ydc}
So for $0.02\leq c_s\leq 1$,  $(c_{s})^{[\epsilon_{V}/2(\epsilon_{V}-1)]}\approx 1$ in the Eq(5.21). So the contribution in the scale of $M_{s}$ comes from $r_{*}$ and the prefactor sitting in the above Eq.~(\ref{hscalecon1}).}.
Additionally, we also obtain a lower bound on $M_s$ 
by considering the lower bound of the sound speed at, $c_s=0.02$, which will generate very small value of 
the tensor-to-scalar ratio $r_{\star}\sim {\cal O}(10^{-22})$ and satisfies the Planck observational constraints. Consequently we 
get the lower bound of the scale of the heavy field at, $M_s\sim {\cal O}(10^{11})$~GeV.

At this point one might worry about the large vacuum energy density stored in the heavy field. This indeed helps inflation, in particular ameliorating the fine tuning
parameter, we have taken $\delta\sim 10^{-4}$ in our scans~\cite{Enqvist:2010vd,Mazumdar:2011ih}. However such a large vacuum energy would need to 
be canceled after the end of slow roll inflation. In the string landscape \cite{Douglas:2006es}, or in the MSSM landscape~\cite{Allahverdi:2008bt}, it is plausible 
to have a bubble nucleation provided the rate of nucleation is large than the Hubble expansion rate. In the context of MSSM, these bubbles will naturally 
yield a low energy vacuum which is an {\it enhanced gauge symmetry point}, first suggested in Ref.~\cite{Allahverdi:2008bt}. In the string vacua case, it is a challenge that 
the false vacuum governed by the heavy field $s$ would nucleate to the MSSM vacuum~\cite{Allahverdi:2007wh}. Furthermore, the bubble nucleation could 
lead to an observational effects such as gravitational waves, etc. ~\cite{Maggiore}. One may be able to constrain further the scale of heavy physics, $M_s$, from the
high frequency gravitational waves,  here we will not discuss these issues any further but we will leave this for future investigation. We can also envisage a smooth
phase transition of the false vacuum as it can happen in the case of hybrid inflation~\cite{Linde:1991km,Linde:1993cn}, possibly triggered by the MSSM inflaton itself 
as discussed in Ref.~\cite{Enqvist:2010vd}. In any of these scenarios we do not expect any modification on large scales, and therefore we do not expect these events to 
affect the primordial perturbations.

%%%%%%%%%%%%%%%%%%%%%%%%%%%%%%%%%%%%%%%%%%%%%%%%%%
%%%%%%%%%%%%%%%%%%%%%%%%%%%%%%%%%%%%%%%%%%%%%%%%%%%%%%%%%%%%%%%%%%%%%%%%%%%%%%%%%%%%%%%%%%%%%%%%%%%%%%%%%%%
\section{Conclusion}

In this paper, we have shown that in any ${\cal N}=1$
SUGRA inflation  model when ever there are more degrees of freedom, non-minimal K\"ahler corrections would
induce three distinct types of corrections: (i) non-minimal kinetic term for 
the inflaton, (ii) Hubble-induced mass correction to the inflaton, and 
(iii) Hubble-induced $A$-term in the potential.

The exact nature of K\"ahler potential and K\"ahler corrections might not be known in all possible scenarios, but 
our aim has been to constrain the coefficients of the non-renormalizable K\"ahler higher dimensional operators phenomenologically, which are gauge invariant,
from the recent Planck data. We assumed minimal K\"ahler potentials for all the fields to begin with.
We first considered the heavy physics to be completely 
decoupled from the dynamics of the light inflaton field. We considered the light field to be 
embedded within MSSM, such that the reheating of the universe is guaranteed to be that of the SM dof. In the simplest setup 
when the heavy field is well settled down in its potential, it only affects via its vacuum energy density. The kinetic terms are mostly canonical, and therefore
we do not obtain any constraint on the coefficients of the dimensional $3$ and $4$ non-reormalizable K\"ahler operators.

We further investigated an intriguing possibility, when the heavy field is coherently oscillating with a frequency larger than the Hubble parameter during the onset of inflation, while
the light field is slowly rolling over the potential.  In this particular scenario, we were able to constrain the coefficients of the Planck suppressed K\"ahler operators of dimensional
$3$ and $4$. We scanned the four parameters, $a,~b,~,c~,d$, and obtained a region of the parameter space where we can satisfy the current Planck observations, i.e. 
$P_S,~n_S,~c_s$ and $r_\star$ within $2\sigma$~CL, and we obtained all the coefficients to be of order $a,~b,~c,~d \sim {\cal O}(1)$, as naturally expected in any non-rrenormalizable SUGRA theory. In fact, as we can see from Fig.~(\ref{fz}) their magnitudes are always less than one.

In Fig.~(\ref{fgh}), we have shown for the range of non-renormalizable corrections, the parameter space for the allowed range of $P_S$ versus $n_S$ for the allowed range of 
$0.02\leq c_s\leq 1$.  In Fig.~(\ref{fig2con}),  we have plotted $r_\star$ vs. $n_S$, for $c_s=1$ and $c_s=0.02$ for the number of e-foldings, $N=50,~70$.
For the range of parameter space scanned, we were able to set an upper limit on the scale of new physics from the constraints arising from $r_\star$, which we obtained 
to be within $10^{11}\leq M_s\leq 10^{16}$~GeV. For the lower bound on $M_s$, we found $r_\star\sim {\cal O}(10^{-22})$ and extremely negligible, and for
the upper bound we saturated $r_\star =0.12$. Note that the current Planck data mildly prefers lower value of the speed of sound, i.e. $c_s < 1$, this is visible from our scans 
and the plot on $r_\star$ versus $n_S$, see Fig.~(\ref{fig2con}).

Finally, we would like to mention that all the above bounds have been obtained for a very particular kind of inflation model, which 
is fully embedded within MSSM, the inflaton is an MSSM flat direction and inflation happens at the point of inflection with a fine tuned parameter at the inflection point is roughly one part in $10^{4}$. We chose MSSM inflation for its advantage that the dynamics can be well understood during inflation and after inflation. In particularly, we can ascertain that the universe after inflation would be filled with the SM degrees of freedom, and also the model is capable of explaining the Higgs mass constraint and the dark matter abundance,
along with the constraints on the inflaton mass arising from the LHC~\cite{Allahverdi:2007vy}. Not every model of inflation enjoys such advantages, and therefore studying this model in some details along with SUGRA corrections yielded interesting constraints. Our methodology can 
be followed for other kinds of inflationary models too.

There is a further scope of improvement in our analysis. So far we have only used the Planck constraints from the power spectra, $P_S$, spectral tilt, $n_S$, tensor-to-scalar ratio, 
$r_\star$, and the constraint on the speed of sound, $c_s$. In principle we should be able to use the non-Gaussian parameters, $f_{NL}^{local}$, $g_{NL}^{local}$ and possibly 
$\tau_{NL}^{local}$, to further constrain the non-renormalizable K\"ahler operators of dimension $3$ and $4$. In our companion paper, we would consider the non-Gaussian constraints 
in some details. All these cosmological constraints arising from Planck and future CMB missions can further improve our understanding of many different aspects of physics beyond the SM.
With an improvement on tensor-to-scalar ratio, $r_\star$, we would be able to further constraint the scale of heavy physics, $M_s$.

%{\rm Acknowledgments:}
\section*{Acknowledgments:}
We would like to thank Lingfei Wang for collaborating at the initial stages. 
SC thanks Council of Scientific and
Industrial Research, India for financial support through Senior
Research Fellowship (Grant No. 09/093(0132)/2010). 
SC also thanks Daniel Baumann and Lingfei Wang for the various useful discussions. SC
also thanks The Abdus Salam International Center for
Theoretical Physics,Trieste, Italy, the organizers of
SUSY 2013 conference and 8th Asian School on Strings, Particles and Cosmology, 2014 for the hospitality during the
work. AM is supported 
by the Lancaster-Manchester-Sheffield Consortium for Fundamental Physics under STFC grant ST/J000418/1.

%%%%%%%%%%%%%%%%%%%%%%%%%%%%%%%%%%%%%%%%%%%%%%%%%%%%%%%%%%%%%%%%%%%%%%%%%%%%%%%%%%%%%%%%%%%%%%%%%%%%%%%%%%%%%%%%%%%%%%%%%%%%%%%%%%%%%%%%%%%%%%%%%%%%%%%%%%%%%%%%%%%%%%%%%%%%%%%%%%%%%%%%
%%%%%%%%%%%%%%%%%%%%%%%%%%%%%%%%%%%%%%%%%%%%%%%%%%%%%%%%%%%%%%%%%%%%%%%%%%%%%%%%%%%%%%%%%%%%%%%%%%%%%%%%%%%%%%%%%%%%%%%%%%%%%%%%%%%%%%%%%%%%%%%%%%%%%%%%%%%%%%%%%%%%%%%%%%%%%%%%%%%%%%%%%%%%%%%%%%%
\section*{Appendix}

%\begin{widetext}

\subsection*{A. XYZ}

The symbols appearing in the Eq~(\ref{sigbz}), in the definition of the sound speed $c_s$ for $I<< M_{p}$, after imposing the slow-roll approxiation are given by:
%\begin{widetext}
 \be\begin{array}{lll}\label{de23}
 \displaystyle {\bf X}_{1}(t)= \sqrt{\frac{2\epsilon_{V}(\phi)V(\phi)}{3}}
\left\{ \sqrt{\frac{2\epsilon_{V}(\phi)V(\phi)}{3}}\frac{aM^{3}_{s}}{M^{2}_{p}}\left[2\sin(2M_{s}t)+4\cos(M_{s}t)\right]\right.\\ \left.
\displaystyle~~~~~~~~~~~~~~~~~~~~~~~~~~~~~~~~~~~~~~~~~~~~
-\frac{aM^{4}_{s}}{M^{2}_{p}}\lvert\phi\rvert\cos{\bf\Theta}\left[\cos(2M_{s}t)-\sin(M_{s}t)\right]\right\},\\
\displaystyle {\bf Y}_{1}(t)= \sqrt{\frac{2\epsilon_{V}(\phi)V(\phi)}{3}}\left\{ \sqrt{\frac{2\epsilon_{V}(\phi)V(\phi)}{3}}\frac{2bM^{2}_{s}}{M_{p}}\cos(M_{s}t)
%\right.\\ \left.
%\displaystyle~~~~~~~~~~~~~~~~~~~~~~~~~~~~~~~~~~~~~~~~
+\frac{bM^{3}_{s}}{M_{p}}\lvert\phi\rvert\cos{\bf\Theta}\sin(M_{s}t)\right\},\\
\displaystyle {\bf Z}_{1}(t)= \sqrt{\frac{2\epsilon_{V}(\phi)V(\phi)}{3}}\left\{\sqrt{\frac{2\epsilon_{V}(\phi)V(\phi)}{3}}\frac{cM^{3}_{s}}{4M^{2}_{p}}\left[2\sin(2M_{s}t)+4\cos(M_{s}t)\right]
\right.\\ \left.
\displaystyle~~~~~~~~~~~~~~~~~~~~~~~~~~~~~~~~~~~~~~~~~~~
-\frac{cM^{4}_{s}}{4M^{2}_{p}}\lvert\phi\rvert\cos{\bf\Theta}\left[\cos(2M_{s}t)-\sin(M_{s}t)\right]\right\},\\
\displaystyle {\bf W}_{1}(t)= \sqrt{\frac{2\epsilon_{V}(\phi)V(\phi)}{3}}\left\{ \sqrt{\frac{2\epsilon_{V}(\phi)V(\phi)}{3}}\frac{4dM^{2}_{s}}{M_{p}}\cos(M_{s}t)
%\right.\\ \left.
%\displaystyle~~~~~~~~~~~~~~~~~~~~~~~~~~~~~~~~~~~~~~~~
+\frac{dM^{3}_{s}}{M_{p}}\lvert\phi\rvert\cos{\bf\Theta}\sin(M_{s}t)\right\},\\
\displaystyle {\bf X}_{2}(t)=\left({\bf Y}_{2}(t)+\frac{a|\phi|^{2}M^{5}_{s}}{M^{2}_{p}}\sin(2M_{s}t)\right),\\
\displaystyle {\bf Y}_{2}(t)={\bf Z}_{2}(t)={\bf W}_{2}(t)=5M^{5}_{s}\sin(2M_{s}t)+8M^{5}_{s}\cos(M_{s}t),\\
\displaystyle {\bf X}_{3}(t)=\left({\bf Y}_{3}(t)-\frac{a|\phi|^{2}M^{5}_{s}}{M^{2}_{p}}\sin(2M_{s}t)\right),\\
\displaystyle {\bf Y}_{3}(t)={\bf Z}_{3}(t)={\bf W}_{3}(t)=3M^{5}_{s}\sin(2M_{s}t)-8M^{5}_{s}\cos(M_{s}t).\\
    \end{array}\ee
%\end{widetext}
Here the complex inflaton field $\phi$ is parameterized by, $\phi=\lvert\phi\rvert\exp(i{\bf\Theta})$. Here the new parameter ${\bf\Theta}$ characterizes 
the phase factor associated with the inflaton and it has a two dimensional rotational symmetry.

%%%%%%%%%%%%%%%%%%%%%%%%%%%%%%%%%%%%%%%%%%%%%%%%%%%%%

%\subsection*{Potential}

%%%%%%%%%%%%%%%%%%%%%%%%%%%%%%%%%%%%%%%%%%%%%%%%%%%%%%%

\subsection*{B. Case -1,~2,~3,~4}

\begin{itemize}

\item{$ {\bf Case-1}~~~~~~~~~~K= \phi^\dag \phi+s^\dag s+\frac{a}{M_{p}^2}\phi^\dag \phi s^\dag s $\\

For the above non-minimal K\"ahler interaction with $'a'$ being a dimensionless number.
%the K\"ahler metric and its inverse metric can be computed as:
%\begin{widetext}
%\begin{equation}\label{met}
% K_{ij^*}=\frac{\partial^2K}{\partial\Phi_i\partial\Phi_{j^*}}=\left(
% \begin{matrix}
%  1+\frac{aI^{\dagger}I}{M_{p}^2} & \frac{aI\phi^{\dagger}}{M_{p}^2}  \\
 % \frac{aI^{\dagger}\phi}{M_{p}^2} & 1+\frac{a\phi^{\dagger}\phi}{M_{p}^2}  \\ 
 %\end{matrix}\right)
%\end{equation}
%\end{widetext}
%which satisfies the orthonormalization criteria, $ K_{ij^*}K^{jm^*}=\delta_{i}^{m}$. 
 We have also computed the correction to the Hubble-induced mass term, for $c_{H} $ for $|I|<<M_{p}$:
%\begin{widetext}
\be
\small c_{H}=\left\{ 3 \left[ (1-a) + (1+a)a\frac{|s|^2}{M_{p}^2}\right]+\left[ (1+3a) + (1-3a)a\frac{|s|^2}{M_{p}^2} \right] \left( \frac{e^K|F_s|^2}{V(s)} -1 \right)
\right\}\approx
                    \displaystyle 3(1-a) \,,
\ee
%\end{widetext}
where we used the fact that: $V(s)=|W_s|^2=3H^2 M^2_{p}=4M^{2}_{s}|s|^2$. Next we compute the correction to the Hubble-induced A term, 
$a_{H}H\frac{\phi^n}{nM_{p}^{n-3}}$, in presence of the non-minimal K\"ahler correction:
%\begin{widetext}
\be\begin{array}{lll}\label{a1}
 a_{H}H\frac{\phi^n}{nM_{p}^{n-3}}\\  =\Big(\Big[1+a\frac{|s|^2}{M_{p}^2}\Big]W_\phi \, \phi -3W(\phi)
\Big)\frac{e^K W^*(I^\dagger)}{M_{p}^2} + \left[W(\phi) \frac{I^\dagger}{M_{p}}
-a  W(\phi) \frac{I^\dagger}{M_{p}} \frac{|I|^2}{M_{p}^2}
 \right.\\ \left.~~~~~~~~~~~~~~~~~~~~~~~~~~~~~~- a W_\phi \, \phi \,  \frac{I^\dagger}{M_{p}}
   \left( 1 - a \frac{|s|^2}{M_{p}^2} \right)\right] \displaystyle \frac{e^K F^*_{\bar{s}}}{M_{p}}+h.c. \\
 \approx \left\{\left( 1 + a \frac{|s|^2}{M_{p}^2} \right)\Big(1-\frac{3}{n}\Big)\frac{ s^2}{M^{2}_{p}}
+ \left( 1 - a \frac{|s|^2}{M_{p}^2} \right)\Big(a-\frac{1}{n}\Big) \frac{(s^\dagger)^2}{M^{2}_{p}} \right\}\frac{\lambda M_{s}\phi^n}{M^{n-3}_{p}}+h.c. \,,
\end{array}\ee
%\end{widetext}
which explicitly shows the  Planck suppression for $|s|<<M_{p}$ in the Hubble-induced A term.}\\

\item{$ {\bf Case-2}~~~~~~~~~~~~K=\phi^\dag\phi +s^\dag s+\frac{b}{2M_{p}}s^\dag\phi \phi + h.c.$\\

Similarly, for the above non-minimal k\"ahler correction where $'b'$ is a dimensionless number
%the K\"ahler metric and its inverse metric can be computed as:
%\begin{widetext}
%\be\begin{array}{lll}
%\displaystyle K_{ij^*}=\frac{\partial^2K}{\partial\Phi_i\partial\Phi_{j^*}}=\left(
% \begin{matrix}
%  1 & \frac{b\phi}{2M_{p}}  \\
%  \frac{b\phi^{\dagger}}{2M_{p}} & 1  \\ 
% \end{matrix}\right)
%\end{array}\ee
%\end{widetext}
%which satifies the similar orthonormalization criteria as stated in the previous case. 
%
we can compute the correction to the Hubble-induced mass
term, $c_{H}H^2 |\phi|^2$, for $|s|\ll M_{p}$:
%\begin{widetext}
\be\begin{array}{lll}
\displaystyle c_{H}=\left[ 3+ b^2\frac{e^K|W_s|^2}{H^2 M_{p}^2} + \left( \frac{e^K|F_s|^2}{V(s)} -1 \right)\right]\approx  3(1+b^2)\,,
\end{array}\ee
%\end{widetext}
where  $V(s)=|W_s|^2=3H^2 M^2_{p}=4M^{2}_{s}|s|^2$. And similarly the Hubble-induced A term, $a_{H}H\frac{\phi^n}{nM_{p}^{n-3}}$, in 
presence of snon-minimal K\"ahler correction read as:
%\begin{widetext}
\be\begin{array}{llll}\label{a2}
 a_{H}H\frac{\phi^n}{nM_{p}^{n-3}}\\
=\Big(W_\phi \,\phi \,
- 3 W(\phi) + b W_\phi \, \phi^\dagger  \frac{s}{M_{p}} \Big) \frac{e^K W^*(I^\dagger)}{M^{2}_{p}}-\frac{b}{2} \frac{e^K W^*(I^\dagger)}{M_{p}^2}  
\left( \frac{W_s}{M_{p}}- \frac{s^\dagger}{M_{p}} \frac{W(I)}{M_{p}^2} \right) \phi \phi   
\\ ~~~~~~~~~~~~~~~~~~~~~~~~~~~~~~~~~~
+\Big( W(\phi) \frac{s^\dagger}{M_{p}}-b W_\phi \, \phi^\dagger \Big)\frac{e^K F_{\bar{s}}^*}{M_{p}} +3bH^2 \frac{s^\dagger}{M_{p}} \phi \phi+ h.c.\\
 =\left\{\left(1-\frac{3}{n}\right)\phi +\frac{ b  \phi^\dagger  s}{n M_{p}} \right\} \frac{\lambda \phi^{n-1}M_{s} s^2}{M_{p}^{n-1}}  
%\\  ~~~~~~~~~~~~~~~~~~~~~~~~~~~~~~~~~~~~~
+\Big(\frac{s^\dagger \phi}{M_{p}}-bn\phi^\dagger \Big)\frac{2M_{s}\lambda \phi^{n-1} s^{\dagger}}{n M^{n-2}_{p}} +\frac{4M^{2}_{s} b |s|^2 s^\dagger}{M^3_{p}} \phi \phi\\
 ~~~~~~~~~~~~~~~~~~~~~~~~~~~~~~~~~~~~~~~~~~~~~~~~~~~ -\frac{b M_{s} s^2}{2M_{p}^2}  
\left( \frac{2M_{s} s}{M_{p}}-\frac{M_{s} s^2 s^\dagger}{M_{p}^3} \right) \phi \phi + h.c.
\end{array}\ee}\\
%\end{widetext}

\item{${\bf Case-3}~~~~~~~~~~~~~~K=\phi \phi^\dag+ss^\dag+\frac{c}{4M_{p}^2}s^\dag s^\dag\phi \phi + h.c. $\\

In a similar way we can analyse  the above non-minimal K\"ahler interaction, where $c$ is the dimensionless number. 
%Here we follow the similar computational algorithm as mentioned in the previous subsections. Using the non-minimal interaction the K\"ahler metric and its inverse metric can be computed as:
%\begin{widetext}
%\be\begin{array}{lll}
%\displaystyle K_{ij^*}=\frac{\partial^2K}{\partial\Phi_i\partial\Phi_{j^*}}=\left(
% \begin{matrix}
%  1 & \frac{c I^{\dagger} \phi}{4M^{2}_{p}}  \\
%  \frac{c I \phi^{\dagger}}{4 M^{2}_{p}} & 1  \\ 
% \end{matrix}\right)
%\end{array}\ee
%\end{widetext}
%which satifies the similar orthonormalization criteria as stated in the previous cases. 
We have computed the correction to the Hubble-induced mass
term, $c_{H}H^2 |\phi|^2$ for $|s|\ll M_{p}$ as:
%
%\begin{widetext}
\be\begin{array}{lll}
\displaystyle c_{H}=\left[ 3+ \frac{3c}{2} \frac{|s|^2}{M_P^2} + \left( 1+ \frac{3c}{2}  \frac{|s|^2}{M_P^2} 
- \frac{c^2}{4} \frac{|s|^4}{M_P^4} \right) \left( \frac{e^K|F_s|^2}{V(s)} -1 \right)\right]\approx 
%\left\{
%	\begin{array}{ll}
                    \displaystyle 3 %\mbox{ for $|I|<<M_{p}$}\,,
 %        \displaystyle 3\left(1+\frac{c}{2}\right) & \mbox{ for $|I|\sim M_{p}$} .
  %        \end{array}
%\right. 
\end{array}\ee
%\end{widetext}
where we have used  $V(s)=|W_s|^2=3H^2 M^2_{p}=4M^{2}_{s}|s|^2$. Next we compute  the Hubble-induced A term, 
$a_{H}H\frac{\phi^n}{nM_{p}^{n-3}}$:
%\begin{widetext}
\be\begin{array}{llll}\label{a3}
 a_{H}H\frac{\phi^n}{nM_{p}^{n-3}}\\=\Big(W_\phi \,\phi \,
- 3 W(\phi) + \frac{c}{2} W_\phi \, \phi^\dagger  \frac{ss}{M_{p}^2} - \frac{c}{2} \frac{s^\dagger}{M_{p}}\frac{W_s}{M_{p}} \phi\phi
\Big) \frac{e^K W^*(I^\dagger)}{M^{2}_{p}}  
%\\ \displaystyle ~~~~~~~~
+\Big( W(\phi) \frac{s^\dagger}{M_{p}}- c W_\phi \, \phi^\dagger  \frac{s}{M_{p}} \Big)\frac{e^K F_{\bar{s}}^*}{M_{p}} \\ 
~~~~~~~~~~~~~~~~~~~~~~~~~~~~~~~~~~~~~~~~~~~~~~~~~~~~~~~~~~~~~~~~~~~~~~~~~~~~~~~~~+\frac{3cH^2}{4}
\frac{s^\dagger s^\dagger}{M^{2}_{p}} \phi \phi
+ h.c.\\
=\left\{\left(1-\frac{3}{n}\right)\phi  +\frac{c\phi^\dagger  ss}{2M_{p}^2}\right\} \frac{\lambda \phi^{n-1}M_{s} s^2}{M_{p}^{n-1}}
  -\frac{c M^{2}_{s} I^2I^\dagger I \phi\phi  }{M^4_{p}} 
%\\ \displaystyle ~~~~~~~~
+\Big(\frac{s^\dagger \phi}{M_{p}}-\frac{cn\phi^\dagger s }{M_{p}} \Big)\frac{2M_{s}\lambda \phi^{n-1} s^{\dagger}}{n M^{n-2}_{p}}\\~~~~~~~~~~~~~~~~~~
~~~~~~~~~~~~~~~~~~~~~~~~~~~~~~~~~~~~~~~~~~~~~~~~~~~~~~~~~~~~~~ +\frac{M^{2}_{s} c 
|s|^2 s^\dagger s^\dagger}{M^4_{p}} \phi \phi + h.c.
\end{array}\ee}\\
%\end{widetext}

\item{${\bf Case-4}~~~~~~~~~~~~~~K=\phi \phi^\dag+ss^\dag+\frac{d}{M_{p}}s\phi^\dag \phi + h.c.$\\

For the above non-minimal K\"ahler potential, where $d$ is the dimensionfull number,
%Here we follow the similar computational algorithm as mentioned in the previous subsections. Using the non-minimal interaction the K\"ahler metric and its inverse metric can be computed as:
%\begin{widetext}
%\be\begin{array}{lll}
%\displaystyle K_{ij^*}=\frac{\partial^2K}{\partial\Phi_i\partial\Phi_{j^*}}=\left(
% \begin{matrix}
%  1+\frac{d I }{M_{p}}+\frac{d I^{\dagger} }{M_{p}} & \frac{d \phi^{\dagger}}{M_{p}}  \\
%  \frac{d \phi}{4 M^{2}_{p}} & 1 \\ 
 %\end{matrix}\right)
%\end{array}\ee
%\end{widetext}
%which satifies the similar orthonormalization criteria as stated in the previous cases.
we can compute the Hubble-induced mass
term, $c_{H}H^2 |\phi|^2$, for $|s|\ll M_{p}$:
%\begin{widetext}
\be\begin{array}{lll}
 c_{H}=\left[3 \left[ 1 + d \frac{s+s^\dagger}{M_P} 
+ d^2 \left( 1 + d \frac{s+s^\dagger}{M_P} \right)^{-1} \right] +\left[ 1 + d \frac{s+s^\dagger}{M_P} 
\right.\right.\\ \left.\left.~~~~~~~~~~~~~~~~~~~~~~~~~~~+ 3 d^2 \left( 1 + d \frac{s+s^\dagger}{M_P} \right)^{-1} \right] \left( \frac{e^K|F_s|^2}{V(s)} -1 \right) \right] %\\ \displaystyle 
\approx 
%\left\{
%	\begin{array}{ll}
                    \displaystyle 3(1+d^2) \,,
%         \displaystyle 3\left(1+2d+\frac{d^2}{1+2d}\right) & \mbox{ for $|I|\sim M_{p}$} .
%          \end{array}
%\right. 
\end{array}\ee
%\end{widetext}
where we used  $V(s)=|W_s|^2=3H^2 M^2_{p}=4M^{2}_{s}|s|^2$. Next we compute the correction to the Hubble-induced A term, $a_{H}H\frac{\phi^n}{nM_{p}^{n-3}}$,
%\begin{widetext}
\be\begin{array}{llll}\label{a4}
 a_{H}H\frac{\phi^n}{nM_{p}^{n-3}}\\=\Big(W_\phi \,\phi \,
- 3 W(\phi) 
\Big) \frac{e^K W^*(s^\dagger)}{M^{2}_{p}}  
%\\ \displaystyle ~~~~~~~~
+\Big( W(\phi) \frac{I^\dagger}{M_{p}}- d \, W(\phi)\Big)\frac{e^K F_{\bar{s}}^*}{M_{p}} 
+ h.c.\\
=\left(1-\frac{3}{n}\right) \frac{\lambda \phi^{n}M_{s} s^2}{M_{p}^{n-1}}  
%\\ \displaystyle ~~~~~~~~
+\Big(\frac{s^\dagger}{M_{p}} -d\Big)\frac{2M_{s}\lambda \phi^{n} s^{\dagger}}{n M^{n-2}_{p}} + h.c.
\end{array}\ee}
%\end{widetext}

\end{itemize}

%%%%%%%%%%%%%%%%%%%%%%%%%%%%%%%%%%%%%%%%%%%%%%%%%%%%%%%%%%%%%%%%%%

%%%%%%%%%%%%%%%%%%%%%%%%%%%%%%%%%%%%%%%%%%%%%%%%%

\subsection*{C. Expression for $a_H$}\label{expression for $a_H$}

Using these results in Hubble induced A-term, $a_{H}$ can be computed from Eqs.~(\ref{a1}), Eq~(\ref{a2}), Eq~(\ref{a3}) and Eq~(\ref{a4}) for the four physical situations, the simplified 
expressions turn out be:
%\begin{widetext}
\be\label{signgxxx}
\small a_{H}\sim \left\{
	\begin{array}{ll}
                      \frac{n}{2}\left(\frac{2}{3}\right)^{\frac{3}{4}}\sqrt{\frac{H_{inf}}{M_{p}}}
\left[1+a-\frac{4}{n}+\frac{35a}{4}\sqrt{\frac{2}{3}}\left(2-\frac{3}{n}\right)\frac{H_{inf}}{M_{p}}-\frac{35a^{2}}{4}\sqrt{\frac{2}{3}}\frac{H_{inf}}{M_{p}}\right]
                    \
  & \mbox{ for $\underline{\bf Case ~I}$}  \\ \\
   \frac{1}{2}\left[3\left(1-\frac{1}{n}\right)+\frac{5b}{n}\sqrt[4]{\frac{2}{3}}\sqrt{\frac{H_{inf}}{M_{p}}}\right] \left(\frac{2}{3}\right)^{\frac{3}{4}}\sqrt{\frac{H_{inf}}{M_{p}}}
+2\sqrt{\frac{2}{3}}\left( \left(\frac{3}{2}\right)^{\frac{3}{4}}\sqrt{\frac{H_{inf}}{M_{p}}}-bn\right)
\\ +10b\left(\frac{3}{2}\right)^{\frac{5}{4}}\left(\frac{M_{p}}{\phi}\right)^{n-2}\left(\frac{H_{inf}}{M_{p}}\right)^{\frac{3}{2}}
-\frac{b}{2}\left(\frac{3}{2}\right)^{\frac{5}{4}}\left(\frac{M_{p}}{\phi}\right)^{n-2}\left(\frac{H_{inf}}{M_{p}}\right)^{\frac{3}{2}}
\left(5-\frac{67}{8}\sqrt{\frac{2}{3}}\frac{H_{inf}}{M_{p}}\right)& \mbox{ for $\underline{\bf Case ~II}$}  \\ \\
   \sqrt[4]{\frac{2}{3}}n\left[\sqrt{\frac{3}{2}}\left(1-\frac{3}{n}\right)+\frac{35c}{24}\frac{H_{inf}}{M_{p}}\right]\sqrt{\frac{H_{inf}}{M_{p}}}
+(1-cn)\sqrt[4]{\frac{2}{3}}\sqrt{\frac{H_{inf}}{M_{p}}} & \mbox{ for $\underline{\bf Case ~III}$}  \\ \\
          \sqrt[4]{\frac{2}{3}}(n-3)\sqrt{\frac{H_{inf}}{M_{p}}}+2\sqrt{\frac{2}{3}}\left[\left(\frac{3}{2}\right)^{\frac{3}{4}}\sqrt{\frac{H_{inf}}{M_{p}}}-d\right] & \mbox{ for $\underline{\bf Case ~IV}$}.
          \end{array}
\right.
\ee

%%%%%%%%%%%%%%%%%%%%%%%%%%%%%%%%%%%%%%%%%%%%%%%%%%%%%%%%%%%%%%%%%%

%%%%%%%%%%%%%%%%%%%%%%%%%%%%%%%%%%%%%%%%%%%%%%%%%

%\end{widetext}
%%%%%%%%%%%%%%%%%%%%%%%%%%%%%%%%%%%%%%%%%%%%%%%%

%%%%%%%%%%%%%%%%%%%%%%%%%%%%%%%%%%%%%%%%%%%%%%%%

\end{document}